\documentclass[pre,aps,preprint,amsmath,amssymb]{revtex4}
\usepackage{graphicx}
\usepackage{epsfig}
\include{graphics}
\usepackage{epstopdf} 
\begin{document}

\title{Fast two-pulse collisions in linear diffusion-advection systems  
with weak quadratic loss in spatial dimension 2}

\author{Avner Peleg$^{1}$ and Toan T. Huynh$^{2}$}

\affiliation{$^{1}$ Department of Mathematics, Azrieli College of Engineering, 
Jerusalem 9371207, Israel}
\affiliation{$^{2}$ Department of Mathematics, University of Medicine and Pharmacy 
at Ho Chi Minh City, Ho Chi Minh City, Vietnam}


\begin{abstract} 
We investigate the dynamics of fast two-pulse collisions in linear 
diffusion-advection systems with weak quadratic loss in spatial 
dimension 2. We introduce a two-dimensional perturbation method, 
which generalizes the perturbation method used for studying 
two-pulse collisions in spatial dimension 1. 
We then use the generalized perturbation method to show that a 
fast collision in spatial dimension 2 leads to a change in the 
pulse shape in the direction transverse to the advection velocity vector. 
Moreover, we show that in the important case of a separable initial 
condition, the longitudinal part in the expression for the amplitude shift 
has a simple universal form, while the transverse part does not.  
Additionally, we show that anisotropy in the initial condition 
leads to a complex dependence of the amplitude shift on 
the orientation angle between the pulses.     
Our perturbation theory predictions are in very good agreement 
with results of extensive numerical simulations with the 
weakly perturbed diffusion-advection model. 
Thus, our study significantly enhances and generalizes the results 
of previous works on fast collisions in diffusion-advection systems, 
which were limited to spatial dimension 1.

\end{abstract}

\maketitle

\section{Introduction}
\label{Introduction}

Linear evolution models have a major role in many areas of physics. 
A central example is provided by the linear diffusion 
equation \cite{Van_Kampen}. Other important examples include 
the linear wave equation \cite{Whitham99}, the linear 
Schr\"odinger equation \cite{Merzbacher98}, and the linear 
propagation equation \cite{Ishimaru2017,Siegman86,Kogelnik66}.  
Many of the physical systems that are described by linear 
evolution models are also affected by weak nonlinear 
dissipation \cite{Van_Kampen,Agrawal2007a}, and are therefore  
more accurately described by perturbed linear evolution models 
with weak nonlinear dissipation. In diffusion-advection physical 
systems, which are the subject of the current paper, the nonlinear 
dissipation arises due to chemical reactions \cite{Van_Kampen,Erdi89}.  
The presence of nonlinear dissipation leads to 
new physical effects that do not exist in the unperturbed 
linear physical systems. The change in pulse concentration 
during collisions between pulses of the linear diffusion-advection 
equation is  an important example for such effects \cite{PNH2017B,NHP2021}. 
Since the pulses of the linear physical systems and also of 
their weakly perturbed counterparts are not shape preserving, 
simple dynamics is not expected in these collisions. 
Consequently, one also does not expect to be able to make simple 
general statements about the physical effects of these collisions.

In two previous works \cite{PNH2017B,NHP2021}, we demonstrated that 
the opposite is in fact true in fast two-pulse collisions 
\cite{fast_collisions}. More specifically, we showed that the amplitude 
shifts in fast two-pulse collisions in linear one-dimensional 
physical systems with weak nonlinear dissipation exhibit simple 
soliton-like universal behavior. The behavior was demonstrated for 
two important cases: (1) systems described by the linear diffusion-advection 
equation with weak quadratic loss; (2) systems described by the 
linear propagation equation with weak cubic loss. 
We first developed perturbation methods  
for analyzing the dynamics of fast collisions in these weakly perturbed 
linear physical systems in spatial dimension 1. 
We then used the perturbation methods to show that in 
both systems, the expressions for the collision-induced 
amplitude shifts due to the nonlinear loss have the 
same simple universal form as the expression for the amplitude shift in a fast 
collision between two solitons of the nonlinear Schr\"odinger equation 
in the presence of weak cubic loss. In addition, we found that within 
the leading order of the perturbation theories, the pulse shapes are not 
changed by the collision. The predictions of the perturbation theories  
for the collision-induced amplitude shifts were validated by extensive 
numerical simulations with the two perturbed linear evolution models 
for a variety of initial pulse shapes \cite{PNH2017B,NHP2021}. 
In addition, in Ref. \cite{QMN2018}, we showed that the perturbation methods  
of Refs. \cite{PNH2017B,NHP2021} can also be used to calculate the 
amplitude shifts in fast two-pulse collisions in the presence of 
weak high-order nonlinear loss. The results of Refs. \cite{PNH2017B,NHP2021} 
are rather surprising. More specifically, the pulses in the weakly perturbed 
linear systems considered in these works are not shape preserving. 
Therefore, the common belief is that conclusions obtained from analysis       
of fast soliton collisions would {\it not} be applicable to collisions in 
these perturbed linear systems \cite{Tkach97,Agrawal89,Allen99,Agrawal2019,PCG2003}. 
However, in Refs. \cite{PNH2017B,NHP2021}, we showed that in fact, 
the opposite is true.

We point out that fast two-pulse collisions in the presence of weak 
nonlinear effects are very important in multisequence optical 
communication systems \cite{Tkach97,Agrawal97,Gnauck2008}. The rates 
of transmission of information in these systems are significantly 
enhanced by transmitting many pulse sequences through the same optical medium.
The pulses in each sequence propagate with the same group velocity, 
but the group velocities are different for pulses from different sequences. 
Consequently, collisions between pulses from different sequences are very frequent, 
and their cumulative effect can lead to severe transmission degradation 
and to transmission error \cite{Tkach97,Agrawal97}. Most multisequence  
optical communication systems are weakly nonlinear \cite{Tkach97,Agrawal97}, 
and almost all collisions in these systems are fast 
\cite{PNH2017B,Mollenauer2003,Nakazawa2000}. Therefore, the study of 
fast collisions between pulses or beams of the linear propagation equation 
in the presence of weak nonlinearities is very relevant 
to multisequence optical communication links.       
An important question in this area concerns the characterization 
of the effects of a single fast two-pulse collision by explicit formulas. 
Indeed, such characterization enables the evaluation of the cumulative effects 
of many fast collisions without the need to perform a large number of 
numerical simulations or experiments. Furthermore, it enables the design 
of methods for compensating the destructive cumulative effects of the collisions.

One can adopt the concept of multisequence optical communication links 
in other physical systems. In particular, the concept can be implemented 
in diffusion-advection physical systems by transmitting many pulse 
sequences of different substances (different gases) through the same 
gaseous medium. Similar to the optical systems, the unwanted collisional
effects in the chemical communication systems (due to chemical reactions) 
can be reduced by designing them, such that the nonlinearities are weak 
and the collisions are fast. 
Therefore, questions concerning the characterization of the 
effects of a single fast two-pulse collision 
in the presence of weak nonlinearities are of central importance for 
multisequence chemical communication as well. It should be mentioned 
that chemical and molecular communication attracted intensive research efforts 
in the last decades. Examples of problems considered in this area include 
airborne communication between mammals \cite{McClintock78} or between 
plants and bacteria \cite{Ryu2013}, communication mediated 
by pheromones \cite{Agosta92,Wyatt2003,Akyildiz2017}, 
and communications in aquatic systems \cite{Czarnik94,Hansson2000}. 
Furthermore, controlled experiments of chemical communication 
via a gaseous medium with a single chemical and with two chemicals 
were reported in Refs. \cite{Taylor2018,Taylor2020}.        
The latter studies also demonstrated that the chemical communication 
links can indeed be described by diffusion-advection models. 
Thus, the possibility for realizing multisequence chemical 
communication provides further motivation for our 
current theoretical study.

The three studies in Refs. \cite{PNH2017B,NHP2021} and \cite{QMN2018} 
were limited to spatial dimension 1, and therefore, did not consider 
important effects, which exist only in the high-dimensional 
collision problem. More specifically, it is not clear if the simple 
universal form of the expressions for the collision-induced amplitude 
shifts, which was found in Refs. \cite{PNH2017B,NHP2021} in the one-dimensional 
problem, remains valid in spatial dimension higher than 1. 
It is also unclear if the pulse shapes remain intact in the collision  
in the high-dimensional problem. Furthermore, high-dimensional effects 
due to anisotropy were not considered in Refs. \cite{PNH2017B,NHP2021}. 
Thus, all the key aspects of the fast collision problem in 
spatial dimension higher than 1 were not addressed in previous studies.      
We note that the extension of the perturbation methods to dimension 
higher than 1 is very challenging due to the following reasons. 
First, it is unclear how to extend the perturbation methods   
in a self-consistent way, which preserves the true small parameters 
in the problem. Second, it is also unclear how to simplify the perturbative 
calculations in a manner, which allows one to obtain explicit expressions 
for the collision-induced changes in the pulse shape and amplitude. 
This is particularly true for pulses of the linear diffusion-advection 
equation, since these pulses are not shape preserving.

In the current work, we treat the aforementioned key aspects 
of the high-dimensional fast collision problem. For this purpose, 
we first develop a perturbation method, which generalizes the 
methods that were introduced in Refs. \cite{PNH2017B,NHP2021} 
for the one-dimensional problem in three important ways. 
First, it extends the perturbative calculation 
from spatial dimension 1 to spatial dimension 2, and enables 
further extension of the calculation to a general spatial dimension. 
Second, it provides a calculation of the 
collision-induced dynamics of the pulse shape both inside 
and outside of the collision interval. In this way, it enables 
an accurate comparison between perturbation theory predictions 
and numerical simulations results for the collision-induced change 
in the pulse shape. In contrast, the perturbative 
calculation of Refs. \cite{PNH2017B,NHP2021} was restricted to 
the collision interval only. Third, it enables the discovery 
and characterization of several major collisional effects, 
which exist only in the high-dimensional problem. 
A key component in the generalization of the perturbation 
methods is the application of a rotation transformation, 
such that in the new coordinate system, the advection velocity 
vector (the relative velocity vector between the colliding pulses) is on the $x$-axis. 
This important step enables one to preserve the true small parameters 
in the problem and to obtain explicit expressions for the collision-induced 
changes in the pulse shape and amplitude.

We use the generalized perturbation theory to obtain 
formulas for the collision-induced changes in the pulse shapes 
and amplitudes in spatial dimension 2. 
We find that for a general initial condition, the collision 
leads to a change in the pulse shape in the direction transverse 
to the advection velocity vector. We then study the important 
case of a separable initial condition, and show that in this 
case, the pulse shape in the longitudinal direction is not 
changed by the collision in the leading order of the perturbation 
theory. Moreover, we show that for a separable initial condition, 
the longitudinal part in the expression for the amplitude shift 
is universal, a property that can be useful for the design of 
scalable multisequence chemical communication links. Additionally, 
the transverse part in the expression for the amplitude shift 
is not universal and is proportional to the integral of the product of 
the pulse concentrations with respect to the transverse coordinate. 
We also show that anisotropy in the initial condition 
leads to a complex dependence of the expression for the amplitude 
shift on the orientation angle between the colliding pulses.     
We attribute this complex dependence to the nonseparable 
character of the initial condition in the anisotropic case.

All the predictions of our perturbation theory are in very good agreement 
with results of extensive numerical simulations with the weakly 
perturbed diffusion-advection model. Therefore, our study significantly 
enhances and extends the results of the previous works in Refs. 
\cite{PNH2017B,NHP2021} on fast two-pulse collisions in linear 
diffusion-advection systems with weak quadratic loss, 
which were limited to spatial dimension 1. 
It is interesting to note that we have recently found that similar 
high-dimensional effects exist in fast collisions between 
two optical beams in a linear optical medium with weak cubic loss \cite{PHN2021}. 
We point out that detailed analytic results on collisions 
between pulse solutions of linear or nonlinear evolution models 
in the presence of nonlinear dissipation in spatial dimension 
higher than 1 are scarce, and this is especially true for pulses that 
are not shape preserving. Therefore, the current work and our recent 
work in Ref. \cite{PHN2021} also significantly enhance 
the understanding of the more general high-dimensional problem 
of fast two-pulse collisions in the presence of nonlinear dissipation.

The remainder of the paper is organized as follows. 
In Sec. \ref{da_2D}, we present the generalized perturbation method 
for calculating the dynamics of the pulse shape and amplitude in fast 
collisions between pulses of the linear diffusion-advection model 
in spatial dimension 2. In Sec. \ref{simu}, we present the predictions 
of our perturbation theory and the results of numerical simulations with 
the weakly perturbed linear diffusion-advection model for three important  
collision setups, which demonstrate three major collisional effects 
that exist only in spatial dimension higher than 1.  
In Sec. \ref{discussion}, we discuss the importance of collision-induced 
effects due to quadratic loss in different setups of diffusion-advection 
systems. Section \ref{conclusions} is reserved for our conclusions.   
The four Appendixes contain derivations that support 
the material in Secs. \ref{da_2D} and \ref{simu}.

\section{The perturbation theory for fast two-pulse collisions 
in spatial dimension 2}
\label{da_2D}

\subsection{Introduction}
\label{da_model}

We study the dynamics of a fast collision between pulses of two substances 
(two gases), denoted by 1 and 2, that move in the same gaseous medium 
with different velocities $\mathbf{v_{1}}$ and $\mathbf{v_{2}}$, and 
evolve in the presence of linear diffusion and weak quadratic loss in 
spatial dimension 2. We assume that the two pulses are injected into the 
gaseous medium with these velocities. Since the gaseous medium is dilute 
and since the collision is fast, the velocities $\mathbf{v_{1}}$ and $\mathbf{v_{2}}$ 
remain constant during the collision. We investigate the dynamics of the collision 
by working in the reference frame that is moving together with pulse 1. 
In this reference frame, pulse 1 is at rest, and pulse 2 is moving 
with velocity that is given by the relative velocity vector 
$\mathbf{v_{d}}=\mathbf{v_{2}}-\mathbf{v_{1}}$ (the advection velocity vector).  
This treatment is justified, since the evolution model that describes the 
collision dynamics  is invariant under Galilean transformations.   
For simplicity and without loss of generality, 
we assume that the advection velocity vector (the relative velocity vector) 
lies on the $x$ axis. Equivalently, we can say that 
we apply a rotation transformation from the original coordinate system 
to a new coordinate system, in which the advection velocity vector is on the $x$ axis, 
and that we describe the collision dynamics in the new coordinate system.  
In Appendix \ref{appendA}, we show that the value of the collision-induced 
amplitude shift is invariant under rotation transformations in the $xy$ plane. 
As a result, the choice of the advection velocity vector along the $x$ 
axis does not change the value of the amplitude shift. 
The dynamics of the fast collision is therefore described 
by the following perturbed $(2+1)$-dimensional linear diffusion-advection model:     
\begin{eqnarray} &&
\!\!\!\!\!\!\!\!\!\!\!\!\!\!
\partial _{t}u_{1}=\partial _{x}^{2}u_{1}+\partial _{y}^{2}u_{1}
-\epsilon_{2}u_{1}^{2}-2\epsilon_{2}u_{2}u_{1},
\nonumber \\&&
\!\!\!\!\!\!\!\!\!\!\!\!\!\!
\partial _{t}u_{2}=\partial _{x}^{2}u_{2}+\partial _{y}^{2}u_{2}
-v_{d1}\partial _{x}u_{2}
-\epsilon_{2}u_{2}^{2}-2\epsilon_{2}u_{1}u_{2}.
\label{da1}
\end{eqnarray}    
In Eq. (\ref{da1}), $u_{1}$ and $u_{2}$ are the concentrations of substances 1 and 2,  
$t$ is time, $x$ and $y$ are spatial coordinates, 
$v_{d1}$ is the advection velocity (the velocity difference between the two pulses), 
and $\epsilon_{2}$ is the quadratic loss coefficient, which satisfies $0< \epsilon_{2} \ll 1$ 
\cite{Dimensions1}. The term $-v_{d1}\partial _{x}u_{2}$ in Eq. (\ref{da1}) is 
an advection term due to the velocity difference between the two pulses.  
The terms $-\epsilon_{2}u_{j}^{2}$ and $-2\epsilon_{2}u_{k}u_{j}$ describe 
intra-substance and inter-substance effects due to quadratic loss, respectively.  
As discussed in the following paragraph, these terms are associated with  
chemical reactions. Note that Eq. (\ref{da1}) does not include linear loss terms.  
However, the inclusion of these terms does not alter the form of the expressions 
for the collision-induced changes in pulse shapes and amplitudes. 
Moreover, the effects of linear loss on amplitude dynamics 
can be incorporated in the analysis in the same manner as was done 
in Refs. \cite{PNH2017B,NHP2021} for spatial dimension 1.

The quadratic loss terms in Eq. (\ref{da1}) are associated with 
chemical reactions and can be derived from the kinetic law of mass action \cite{Erdi89,Jameson2001,Tomlin2014,Voit2015}. More specifically, 
the terms $-\epsilon_{2}u_{1}^{2}$ and $-\epsilon_{2}u_{2}^{2}$ 
are due to the reactions $A+A \rightarrow A_{2}$ and $B+B \rightarrow B_{2}$, 
respectively. Additionally, the terms $-2\epsilon_{2}u_{k}u_{j}$ are due to 
the reaction $A+B \rightarrow AB$. We assume that the reactions products 
$AB$, $A_{2}$, and $B_{2}$ do not react with the reactants $A$ and $B$, 
and do not dissociate, i.e., the reactions are irreversible. 
We remark that in Eq. (\ref{da1}), we consider as an example the case where the 
rate coefficients for the reactions $A+A \rightarrow A_{2}$ and $B+B \rightarrow B_{2}$  
are equal to each other, and are also equal to 1/2 of the rate coefficient 
for the reaction $A+B \rightarrow AB$. We emphasize, however, that results similar 
to the ones presented in our paper are obtained in the more general case, where 
the rate coefficients for the reactions do not follow these simple relations. 
We also point out that in general, the reaction $A+B \rightarrow AB$ is independent 
of the reactions $A+A \rightarrow A_{2}$, and $B+B \rightarrow B_{2}$. Therefore, we can 
also consider physical setups where only the rate of the reaction $A+B \rightarrow AB$ is significant, 
while the rates of the reactions $A+A \rightarrow A_{2}$ and $B+B \rightarrow B_{2}$ are negligible. 
In these physical setups, which are considered in section III D, we can neglect the terms 
proportional to $-u_{j}^{2}$ in Eq. (\ref{da1}).

We consider fast collisions between pulses with generic 
initial shapes and with tails that decay sufficiently fast, 
such that the initial values of the total mass 
$\int_{-\infty}^{\infty} dx \int_{-\infty}^{\infty} dy u_{j}(x,y,0)$ 
are finite. We assume that the initial pulses can be characterized by 
initial amplitudes $A_{j}(0)$, initial widths along the $x$ and $y$ axes 
$W_{j0}^{(x)}$ and $W_{j0}^{(y)}$, and initial positions $(x_{j0}, y_{j0})$.  
Thus, the initial concentrations can be written as: 
\begin{eqnarray} &&
u_{j}(x,y,0)=A_{j}(0)h_{j}(x,y),  
\label{da2}
\end{eqnarray} 
where $h_{j}(x,y)$ depends on the parameters $W_{j0}^{(x)}$, $W_{j0}^{(y)}$, 
$x_{j0}$, and $y_{j0}$. 
We are also interested in the important case, where the initial 
concentrations of both substances are separable, i.e., where each of 
the functions $u_{j}(x,y,0)$ can be expressed as a product 
of a function of $x$ and a function of $y$: 
\begin{eqnarray} &&
\!\!\!\!\!\!
u_{j}(x,y,0)=A_{j}(0)h_{j}^{(x)}[(x-x_{j0})/W_{j0}^{(x)}]
h_{j}^{(y)}[(y-y_{j0})/W_{j0}^{(y)}]. 
\label{da3}
\end{eqnarray}       
The separable initial condition (\ref{da3}) is important, 
since in this case one can further simplify the expressions 
for the collision-induced changes in pulse shapes and amplitudes 
and in this manner, gain deeper insight into the collision dynamics. 
In what follows, we will also consider cases where the initial concentration 
is separable for one substance and nonseparable for the other substance.

In the current paper, we are interested in the dynamics of complete 
fast collisions. The complete collision assumption means that the 
pulses are well-separated before and after the collision. 
Therefore, in these collisions, the values of the $x$ 
coordinate of the pulse locations at $t=0$ and at the 
final time $t_{f}$, $x_{j0}$ and $x_{j}(t_f)$, satisfy 
$|x_{20}-x_{10}| \gg W_{10}^{(x)}+W_{20}^{(x)}$ and 
$|x_{2}(t_f)-x_{1}(t_f)| \gg W_{1}^{(x)}(t_f)+W_{2}^{(x)}(t_f)$, 
where $W_{j}^{(x)}(t_f)$ are the pulse widths in the $x$ direction at $t=t_{f}$.  
The assumption of a fast collision means that the collision time interval 
$\Delta t_{c}=2(W_{10}^{(x)}+W_{20}^{(x)})/|v_{d1}|$, which is the time interval 
during which the colliding pulses overlap, is much shorter than the smallest 
diffusion time in the problem. We note that the diffusion times 
for the $j$th pulse in the $x$ and $y$ directions are 
$t_{Dj}^{(x)}=W_{j0}^{(x)2}$ and $t_{Dj}^{(y)}=W_{j0}^{(y)2}$,  
respectively. Therefore, the smallest diffusion time is 
$t_{D}^{(min)}=\min \left\{ t_{D1}^{(x)}, t_{D2}^{(x)}, 
t_{D1}^{(y)}, t_{D2}^{(y)}\right\}$.
Requiring that $\Delta t_{c} \ll t_{D}^{(min)}$, we find that the condition for a fast 
collision is $2(W_{10}^{(x)}+W_{20}^{(x)}) \ll |v_{d1}| t_{D}^{(min)}$.

Let us demonstrate that the condition for a fast collision can be realized 
in many weakly perturbed diffusion-advection systems. For simplicity, 
we look at the case where $W_{10}^{(x)}=W_{10}^{(y)}=W_{20}^{(x)}=W_{20}^{(y)}=W_{0}$,  
such that the fast collision condition takes the form $W_{0}|v_{d1}|/4 \gg 1$. 
Consider for instance diffusion of $\mbox{O}_{2}$ in $\mbox{N}_{2}$ 
at 293.15$^{\circ}$K and at a pressure of 1 atm. 
The diffusion coefficient is $D=0.202$ $\mbox{cm}^{2}/\mbox{s}$ \cite{CRC2004}.  
Thus, for an initial pulse width of 4 cm and for an advection velocity value 
of $V_{d1}=5$ $\mbox{cm}/\mbox{s}$, we find that $W_{0}|v_{d1}|/4 = 24.75$. 
Similar results are obtained for other gas pairs. For example, 
the diffusion coefficient of $\mbox{CO}_{2}$ in $\mbox{O}_{2}$ at 293.15$^{\circ}$K 
and at a pressure of 1 atm is $D=0.159$ $\mbox{cm}^{2}/\mbox{s}$ \cite{CRC2004}. 
Using this value we find that for an initial pulse width of 2 cm and for 
$V_{d1}=5$ $\mbox{cm}/\mbox{s}$, $W_{0}|v_{d1}|/4 =15.72$. Therefore, the 
condition for a fast collision is clearly satisfied in both cases.

\subsection{Calculation of the collision-induced effects for a general initial condition} 
\label{da_general_IC}

\subsubsection{Introduction}     
We introduce a perturbation method, which generalizes the perturbative  
calculation presented in Refs. \cite{PNH2017B,NHP2021} in three important aspects. 
(1) It extends the calculation from spatial dimension 1 
to spatial dimension 2, and enables further extension of the 
calculation to a general spatial dimension. 
(2) It provides a perturbative calculation of the collision-induced change 
in the pulse shape both in the collision interval and away from the collision interval, 
whereas the calculation in Refs. \cite{PNH2017B,NHP2021} was limited to the 
collision interval only. This improvement enables an accurate comparison 
between perturbation theory predictions and numerical simulations results 
for the collision-induced change in the pulse shape. 
(3) It helps uncover several collision-induced 
effects, which exist only in spatial dimension higher than 1, 
and in this manner, enables deeper insight into the collision dynamics 
in the two-dimensional problem.

In the first step in the perturbative calculation, we look for a solution 
of Eq. (\ref{da1}) in the form: 
\begin{eqnarray}&&
\!\!\!\!\!\!\!
u_{j}(x,y,t)=u_{j0}(x,y,t)+\phi_{j}(x,y,t), 
\label{da5}
\end{eqnarray}    
where $j=1,2$, $u_{j0}$ are solutions of Eq. (\ref{da1}) 
without inter-pulse interaction, and $\phi_{j}$ describe collision-induced effects. 
By definition, the $u_{j0}$ satisfy the following equations:  
\begin{eqnarray}&&
\!\!\!\!\!\!\!
\partial_{t}u_{10}=\partial_{x}^{2}u_{10} + \partial_{y}^{2}u_{10} 
-\epsilon_{2}u_{10}^{2},
\!\!\!\!\!\!\!\!
\label{da6}
\end{eqnarray}         
and 
\begin{eqnarray}&&
\!\!\!\!\!\!\!
\partial _{t}u_{20}=\partial _{x}^{2}u_{20} + \partial_{y}^{2}u_{20}
-v_{d1}\partial _{x}u_{20} - \epsilon_{2}u_{20}^{2}.
\!\!\!\!\!\!\!\!
\label{da7}
\end{eqnarray}  
We can expand the $u_{j0}$ in perturbation series with respect to $\epsilon_{2}$: 
\begin{eqnarray} &&
u_{j0}(x,y,t) = A_{j}(t) \tilde u_{j0}(x,y,t) + \tilde u_{j0}^{(1)}(x,y,t) 
+ \tilde u_{j0}^{(2)}(x,y,t) + ...  , 
\label{da7_add1}  
\end{eqnarray}
where $A_{j}(t)$ are time-dependent amplitudes, 
and $\tilde u_{j0}(x,y,t)$ are the solutions of the unperturbed 
linear diffusion-advection equation with unit amplitude.   
In addition, $\tilde u_{j0}^{(1)}$ and $\tilde u_{j0}^{(2)}$ 
are the $O(\epsilon_{2})$ and $O(\epsilon_{2}^{2})$ corrections 
to the latter solutions, and the $...$ stand for terms of order 
$\epsilon_{2}^{3}$ and higher. It is important to note that the 
terms $\tilde u_{j0}$, $\tilde u_{j0}^{(1)}$, $\tilde u_{j0}^{(2)}$ 
are associated with single-pulse evolution effects and not with 
collision-induced effects. In addition, we expand the $\phi_{j}$ 
in perturbation series with respect to the two small parameters 
$\epsilon_{2}$ and $1/|v_{d1}|$. In the current paper, we are 
interested in the first nonzero term in each of the expansions of the $\phi_{j}$. 
These first nonzero terms in the expansions clearly represent the 
leading-order collision-induced changes in the pulse shapes, 
and we therefore refer to them as the leading-order expressions 
for the $\phi_{j}$.

We substitute the ansatz (\ref{da5}) into Eq. (\ref{da1}) 
and use Eqs. (\ref{da6}) and (\ref{da7}) to obtain equations 
for the $\phi_{j}$. We concentrate on the calculation of $\phi_{1}$, 
as the calculation of $\phi_{2}$ is similar. 
To obtain the leading-order expression for $\phi_{1}$, we neglect 
high-order terms containing products of $\epsilon_{2}$ with 
$\phi_{1}$ or $\phi_{2}$. In addition, in the calculation of 
the leading-order expression for $\phi_{1}$, we keep only the 
leading terms in the expansions (\ref{da7_add1}). 
That is, in this calculation, we approximate the $u_{j0}$ by: 
\begin{eqnarray} &&
u_{j0}(x,y,t) \simeq A_{j}(t) \tilde u_{j0}(x,y,t). 
\label{da10}
\end{eqnarray}    
The substitution and subsequent approximations yield the following 
equation for the leading-order expression for $\phi_{1}$:   
\begin{equation}
\partial_{t}\phi_{1}=\partial_{x}^{2}\phi_{1} + \partial_{y}^{2}\phi_{1} 
-2\epsilon_{2} A_{1}(t)A_{2}(t) \tilde u_{20} \tilde u_{10}. 
\label{da8} 
\end{equation}           
Note that for brevity and simplicity of notation, in Eq. (\ref{da8}), 
we denote the leading-order expression for $\phi_{1}$ by $\phi_{1}$. 
This notation is also used in the remainder of the paper.

In solving the equation for $\phi_{1}$, we distinguish between two 
time intervals, the collision interval and the post-collision interval. 
To define these intervals, we introduce the collision 
time $t_{c}$, which is the time at which the $x$ coordinates of 
the colliding pulses coincide, i.e., $x_{1}(t_{c})=x_{2}(t_{c})$. 
The collision interval is the small time interval 
$t_{c} - \Delta t_{c}/2 \le t \le t_{c} + \Delta t_{c}/2$ centered 
about $t_{c}$, in which the two pulses overlap.  
The post-collision interval is defined as the time interval 
$t > t_{c} + \Delta t_{c}/2$, in which the pulses no longer overlap.

\subsubsection{Collision-induced effects in the collision interval} 

We first find the orders of the different terms in  Eq. (\ref{da8}) 
in the collision interval. Since $\Delta t_{c}$ is of order 
$1/|v_{d1}|$, the term $\partial_{t}\phi_{1}$ is of order $|v_{d1}| \times O(\phi_{1})$.  
The term $-2\epsilon_{2} A_{1}(t)A_{2}(t) \tilde u_{20} \tilde u_{10}$  
is of order $\epsilon_{2}$. Equating the orders of $\partial_{t}\phi_{1}$ and 
$-2\epsilon_{2} A_{1}(t)A_{2}(t) \tilde u_{20} \tilde u_{10}$, 
we find that $\phi_{1}$ is of order $\epsilon_{2}/|v_{d1}|$. In addition, 
$\phi_{1}$ does not contain any fast dependence on $x$ and $y$. 
Therefore, the terms $\partial_{x}^{2}\phi_{1}$ and $\partial_{y}^{2}\phi_{1}$   
are of order $\epsilon_{2}/|v_{d1}|$ and can be neglected.   
It follows that the equation for the leading-order expression for  
$\phi_{1}$ in the collision interval is    
\begin{equation}
\partial _{t}\phi_{1}=-2\epsilon_{2}A_{1}(t)A_{2}(t) \tilde u_{20} \tilde u_{10}.
\label{da9} 
\end{equation}             
Equation (\ref{da9}) has the same form as the equation obtained 
for a fast collision between two pulses of the linear diffusion-advection equation 
in the presence of weak quadratic loss in spatial dimension 1 \cite{PNH2017B,NHP2021}.
It is also similar to the equation obtained for a fast collision between two 
solitons of the cubic nonlinear Schr\"odinger equation in the presence 
of weak cubic loss in spatial dimension 1 \cite{PNC2010}.

A similar calculation shows that $\phi_{2}$ is of order $\epsilon_{2}/|v_{d1}|$. 
Therefore, using Eqs. (\ref{da5}) and (\ref{da7_add1}), we obtain that the 
expansions of the total concentrations $u_{j}(x,y,t)$ up to second-order in 
the two small parameters $\epsilon_{2}$ and $1/|v_{d1}|$ are: 
\begin{eqnarray}&&
u_{j}(x,y,t) \simeq 
A_{j}(t) \tilde u_{j0}(x,y,t) + \tilde u_{j0}^{(1)}(x,y,t) 
+ \tilde u_{j0}^{(2)}(x,y,t) + \phi_{j}(x,y,t). 
\label{da11}
\end{eqnarray}  
In addition, the dynamics of the $A_{j}(t)$ that is associated with 
single-pulse evolution is described in Appendix \ref{appendB}.

We calculate the collision-induced amplitude shift of pulse 1 from  
the collision-induced change in the concentration of pulse 1,  
$\Delta\phi_{1}(x,y,t_{c})=\phi_{1}(x,y,t_{c}+\Delta t_{c}/2)-
\phi_{1}(x,y,t_{c}-\Delta t_{c}/2)$. $\Delta\phi_{1}(x,y,t_{c})$ 
is calculated by integrating Eq. (\ref{da9}) with respect to time 
over the collision interval. This calculation yields:  
\begin{eqnarray} &&
\!\!\!\!\!\!\!\!\!\!\!
\Delta\phi_{1}(x,y,t_{c})\!=\!
-2\epsilon_{2}\!\!\int_{t_{c}-\Delta t_{c}/2}^{t_{c}+\Delta t_{c}/2} 
\!\!\!\!\!\!\!\!\!\! dt' A_{1}(t') A_{2}(t')
\tilde u_{10}(x,y,t') \tilde u_{20}(x,y,t'). 
\label{da12}
\end{eqnarray}    
$\tilde u_{20}$ is the only function in the integrand in 
Eq. (\ref{da12}) that contains fast variations in $t$, 
which are of order 1. Therefore, we can approximate $A_{1}(t)$, $A_{2}(t)$, 
and $\tilde u_{10}(x,y,t)$ by $A_{1}(t_{c}^{-})$, $A_{2}(t_{c}^{-})$, 
and $\tilde u_{10}(x,y,t_{c})$, where $A_{j}(t_{c}^{-})$ 
is the limit from the left of $A_{j}$ at $t_{c}$. 
Furthermore, in the calculation of the integral, 
we can take into account in an exact manner only the fast 
dependence of $\tilde u_{20}$ on $t$, i.e., the dependence on $t$ 
that is contained in the factors $\tilde x=x-x_{20}-v_{d1}t$, 
and replace $t$ by $t_{c}$ everywhere else in the expression 
for $\tilde u_{20}$. We denote this approximation of 
$\tilde u_{20}(x,y,t)$ by $\bar u_{20}(\tilde x,y,t_{c})$. 
Carrying out all these approximations in Eq. (\ref{da12}), we obtain:  
\begin{eqnarray} &&
\!\!\!\!\!\!\!\!\!\!\!
\Delta\phi_{1}(x,y,t_{c})\!=\!
-2\epsilon_{2}A_{1}(t_{c}^{-})A_{2}(t_{c}^{-})\tilde u_{10}(x,y,t_{c}) 
\nonumber \\&&
\times
\!\!\!\int_{t_{c}-\Delta t_{c}/2}^{t_{c}+\Delta t_{c}/2} 
\!\!\!\!\!\!\!\! dt'  \bar u_{20}(x-x_{20}-v_{d1}t',y,t_{c}). 
\label{da13}
\end{eqnarray}  
We assume that the integrand on the right hand side 
of Eq. (\ref{da13}) is sharply peaked in a small interval around $t_{c}$.  
As a result, we can extend the integral limits  to $-\infty$ and $\infty$. 
We also change the integration variable from $t'$ to 
$\tilde x=x-x_{20}-v_{d1}t'$ and obtain: 
\begin{eqnarray} &&
\!\!\!\!\!\! 
\Delta\phi_{1}(x,y,t_c)=
-\frac{2\epsilon_{2}A_{1}(t_{c}^{-})A_{2}(t_{c}^{-})}{|v_{d1}|}\tilde u_{10}(x,y,t_{c}) 
\!\int_{-\infty}^{\infty} \!\!\!\!\!\! d\tilde x \, \bar u_{20}(\tilde x,y,t_{c}).
\label{da14}
\end{eqnarray}    
We observe that the $y$ dependence of $\Delta\phi_{1}(x,y,t_{c})$ is 
affected by the $y$ dependence of pulse 2 at $t=t_{c}$, whereas the 
$x$ dependence of $\Delta\phi_{1}(x,y,t_{c})$ is not affected by 
the $x$ dependence of pulse 2. Therefore, inside the collision interval, 
the pulse shape in the longitudinal direction is retained, 
while the pulse shape in the transverse direction is changed. 
The latter collision-induced change is an effect that exists only 
in spatial dimension higher than 1.

One can show that the relation between the collision-induced amplitude 
shift of pulse 1, $\Delta A_{1}^{(c)}$, and $\Delta\phi_{1}(x,y,t_{c})$ 
is: 
\begin{eqnarray}&&
\!\!\!\!\!\!\!\!\!\!\!\!\!\!
\Delta A_{1}^{(c)}= C_{d1}^{-1}
\!\int_{-\infty}^{\infty} \!\!\!\!\! dx 
\!\int_{-\infty}^{\infty} \!\!\!\!\! dy
\, \Delta\phi_{1}(x,y,t_c), 
\label{da15}
\end{eqnarray}        
where 
\begin{eqnarray}&&
C_{d1}= 
\!\!\int_{-\infty}^{\infty} \!\!\!\! dx 
\!\int_{-\infty}^{\infty} \!\!\!\! dy
\;\tilde u_{10}(x,y,0).  
\label{da16}
\end{eqnarray}       
Substitution of Eq. (\ref{da14}) into Eq. (\ref{da15}) yields the following   
expression for $\Delta A_{1}^{(c)}$ for the general initial condition (\ref{da2}): 
\begin{eqnarray} &&
\!\!\!\!\!\!\!\!
\Delta A_{1}^{(c)}=-\frac{2\epsilon_{2}A_{1}(t_{c}^{-})A_{2}(t_{c}^{-})}
{C_{d1}|v_{d1}|}
\!\!\int_{-\infty}^{\infty} \!\!\!\!\! dx 
\!\int_{-\infty}^{\infty} \!\!\!\!\! dy
\;\tilde u_{10}(x,y,t_{c})
\!\int_{-\infty}^{\infty} \!\!\!\!\! d\tilde x \;\bar u_{20}(\tilde x,y,t_{c}). 
\label{da17}
\end{eqnarray}

\subsubsection{Evolution of $\phi_{1}(x,y,t)$ in the post-collision interval}
In the post-collision interval, $t > t_{c} + \Delta t_{c}/2$, 
the colliding pulses are no longer overlapping. As a result, 
in this interval, the term 
$-2\epsilon_{2} A_{1}(t)A_{2}(t) \tilde u_{20} \tilde u_{10}$
in Eq. (\ref{da8}) can be neglected. Therefore, in the leading order 
of the calculation of the collisional effects, the equation describing 
the dynamics of $\phi_{1}(x,y,t)$ in the post-collision interval 
is the unperturbed linear diffusion equation 
\begin{equation}
\partial_{t}\phi_{1}=\partial_{x}^{2}\phi_{1} + \partial_{y}^{2}\phi_{1}. 
\label{da21} 
\end{equation}           
Note that for $|v_{d1}| \gg 1$, $\Delta\phi_{1}(x,y,t_{c}) \simeq 
\phi_{1}(x,y,t_{c}^{+})$, where $\phi_{1}(x,y,t_{c}^{+})$ is the 
limit from the right of $\phi_{1}(x,y,t)$ at $t=t_{c}$.
Therefore, the initial condition for Eq. (\ref{da21}) is: 
\begin{eqnarray}&&
\phi_{1}(x,y,t_{c}^{+}) = \Delta\phi_{1}(x,y,t_{c}),   
\label{da22}
\end{eqnarray}        
where $\Delta\phi_{1}(x,y,t_{c})$ is given by Eq. (\ref{da14}). 
The solution of Eq. (\ref{da21}) with the initial condition (\ref{da22}) is 
\begin{eqnarray}&&
\phi_{1}(x,y,t) = {\cal F}^{-1}\left(\hat\phi_{1}(k_{1},k_{2},t_{c}^{+})
\exp[-(k_{1}^{2} + k_{2}^{2})(t-t_{c})]\right),  
\label{da23}
\end{eqnarray}         
where $\hat\phi_{1}(k_{1},k_{2},t_{c}^{+})=
{\cal F}\left(\phi_{1}(x,y,t_{c}^{+}) \right)$, 
and ${\cal F}$ and ${\cal F}^{-1}$ are the Fourier transform and the 
inverse Fourier transform with respect to $x$ and $y$.

\subsection{Calculation of the collision-induced effects for a separable initial condition} 
\label{da_separable_IC}

\subsubsection{Introduction}  
Let us describe the collision dynamics in the important case, 
where the initial condition is given by Eq. (\ref{da3}), 
i.e., it is separable for both pulses. This case is of special 
interest, since it is possible to further simplify the expressions 
for the collision-induced changes of the pulse shape and amplitude, 
and in this way, obtain deeper insight into the collision dynamics.

The solutions of the unperturbed linear diffusion equation 
with the separable initial condition (\ref{da3}) 
and with unit amplitude can be expressed in the product form: 
\begin{eqnarray}&&
\tilde u_{j0}(x,y,t)=g_{j}^{(x)}(x,t)g_{j}^{(y)}(y,t),
\label{da25}
\end{eqnarray}   
where 
\begin{eqnarray}&&
g_{1}^{(x)}(x,t) = (2\pi)^{-1/2}
\!\int_{-\infty}^{\infty} \!\!\!\! d k_{1} \hat f_{1}^{(x)}(k_{1})
\exp[-k_{1}^{2}t + ik_{1}x],  
\label{da26}
\end{eqnarray}  
\begin{eqnarray}&&
g_{2}^{(x)}(x,t) = (2\pi)^{-1/2}
\!\int_{-\infty}^{\infty} \!\!\!\! d k_{1} \hat f_{2}^{(x)}(k_{1})
\exp[- iv_{d1}k_{1}t - k_{1}^{2}t + ik_{1}x], 
\label{da27}
\end{eqnarray}   
and 
\begin{eqnarray}&& 
g_{j}^{(y)}(y,t) = (2\pi)^{-1/2}
\!\int_{-\infty}^{\infty} \!\!\!\! d k_{2} \hat f_{j}^{(y)}(k_{2})
\exp[-k_{2}^{2}t + ik_{2}y] .
\label{da28}
\end{eqnarray}         
The functions $\hat f_{j}^{(x)}(k_{1})$ and $\hat f_{j}^{(y)}(k_{2})$ in 
Eqs. (\ref{da26})-(\ref{da28}) are the Fourier transforms of 
$h_{j}^{(x)}[(x-x_{j0})/W_{j0}^{(x)}]$ and 
$h_{j}^{(y)}[(y-y_{j0})/W_{j0}^{(y)}]$, respectively. 
In addition, using the conservation of the total mass for the 
unperturbed linear diffusion equation along with Eqs. (\ref{da3}) 
and (\ref{da25}), we obtain  
\begin{eqnarray}&&
\!\!\!\!
\int_{-\infty}^{\infty} \!\!\!\! dx \, g_{j}^{(x)}(x,t)
=\int_{-\infty}^{\infty} \!\!\!\! dx \, g_{j}^{(x)}(x,0)
=W_{j0}^{(x)}\int_{-\infty}^{\infty} \!\!\!\! ds \, h_{j}^{(x)}(s)
=W_{j0}^{(x)}  c_{dj}^{(x)},  
\label{da29}
\end{eqnarray}     
and 
\begin{eqnarray}&&
\!\!\!\!\!\!
\int_{-\infty}^{\infty} \!\!\!\! dy \, g_{j}^{(y)}(y,t)
=\int_{-\infty}^{\infty} \!\!\!\! dy \, g_{j}^{(y)}(y,0)
=W_{j0}^{(y)}\int_{-\infty}^{\infty} \!\!\!\! ds \, h_{j}^{(y)}(s)
=W_{j0}^{(y)}  c_{dj}^{(y)},  
\label{da30}
\end{eqnarray}                    
where $c_{dj}^{(x)}$ and $c_{dj}^{(y)}$ are constants.

\subsubsection{Collision-induced effects in the collision interval}  

We start by calculating the collision-induced change in the pulse 
shape in the collision-interval for an initial condition that is 
separable for both pulses. Note that from the definition of 
$\bar u_{20}(\tilde x,y,t_{c})$ it follows that 
$\bar u_{20}(\tilde x,y,t_{c})=\tilde u_{20}(x,y,t_{c})$.  
Using this relation together with Eqs. (\ref{da25}) and (\ref{da29}), 
we obtain: 
\begin{eqnarray}&&
\int_{-\infty}^{\infty} \!\!\!\!\! d\tilde x \, \bar u_{20}(\tilde x,y,t_{c})=
c_{d2}^{(x)} W_{20}^{(x)} g_{2}^{(y)}(y,t_{c}).
\label{da31}
\end{eqnarray}     
Substitution of Eq. (\ref{da31}) into Eq. (\ref{da14}) yields: 
\begin{eqnarray} &&
\!\!\!\!\!\! 
\Delta\phi_{1}(x,y,t_c)=
-\frac{2\epsilon_{2}A_{1}(t_{c}^{-})A_{2}(t_{c}^{-})}{|v_{d1}|}
c_{d2}^{(x)} W_{20}^{(x)} g_{2}^{(y)}(y,t_{c})
\tilde u_{10}(x,y,t_{c}). 
\label{da32}
\end{eqnarray}     
Equation (\ref{da32}) is valid for an initial condition that is 
separable for pulse 2, but not necessarily separable for pulse 1. 
When the initial condition is separable for pulse 1 as well, we 
obtain: 
\begin{eqnarray} &&
\!\!\!\!\!\! 
\Delta\phi_{1}(x,y,t_c)=
-\frac{2\epsilon_{2}A_{1}(t_{c}^{-})A_{2}(t_{c}^{-})}{|v_{d1}|}
c_{d2}^{(x)} W_{20}^{(x)}g_{1}^{(x)}(x,t_{c})g_{1}^{(y)}(y,t_{c})
g_{2}^{(y)}(y,t_{c}). 
\label{da33}
\end{eqnarray}  
We see that the pulse shape in the longitudinal direction does not 
change inside the collision interval, which is similar to the behavior 
observed for the general initial condition [compare Eqs. (\ref{da14}) 
and (\ref{da33})]. In subsection \ref{dynamics2}, we show that for a 
separable initial condition, the pulse shape in the longitudinal direction 
is not changed at all by the collision (within the leading order of the 
perturbative calculation).

Next, we calculate the collision-induced change in the pulse amplitude 
for a separable initial condition. Using Eqs. (\ref{da16}), (\ref{da3}), 
(\ref{da29}), and (\ref{da30}), we find   
$C_{d1}=c_{d1}^{(x)}c_{d1}^{(y)}W_{10}^{(x)}W_{10}^{(y)}$. 
In addition, using Eqs. (\ref{da25}) and (\ref{da31}), we obtain 
\begin{eqnarray} &&
\!\int_{-\infty}^{\infty} \!\!\!\!\! dx 
\!\int_{-\infty}^{\infty} \!\!\!\!\! dy
\;\tilde u_{10}(x,y,t_{c})
\!\int_{-\infty}^{\infty} \!\!\!\!\! d\tilde x \;\bar u_{20}(\tilde x,y,t_{c})=
\nonumber \\&&
c_{d1}^{(x)}c_{d2}^{(x)}W_{10}^{(x)}W_{20}^{(x)} 
\!\int_{-\infty}^{\infty} \!\!\!\!\! dy \, 
g_{1}^{(y)}(y,t_{c})g_{2}^{(y)}(y,t_{c}). 
\label{da36}
\end{eqnarray}        
Substitution of Eq. (\ref{da36}) and the expression for $C_{d1}$ 
into Eq. (\ref{da17}) yields the following equation for 
$\Delta A_{1}^{(c)}$ for a separable initial condition:  
\begin{eqnarray} &&
\!\!\!\!
\Delta A_{1}^{(c)}=-\frac{2\epsilon_{2}
A_{1}(t_{c}^{-}) A_{2}(t_{c}^{-})}{|v_{d1}|}
\frac{c_{d2}^{(x)}W_{20}^{(x)}} 
{c_{d1}^{(y)}W_{10}^{(y)}}
\!\int_{-\infty}^{\infty} \!\!\!\!\! dy \, 
g_{1}^{(y)}(y,t_{c})g_{2}^{(y)}(y,t_{c}). 
\label{da37}
\end{eqnarray}     
We observe that the expression for $\Delta A_{1}^{(c)}$ has the form 
\begin{eqnarray} &&
\!\!\!\!\!\!\!\!
\Delta A_{1}^{(c)}=-(\mbox{overall factor}) \times 
(\mbox{longitudinal factor}) \times (\mbox{transverse factor}), 
\label{da38}
\end{eqnarray}  
where the overall factor is 
$2\epsilon_{2} A_{1}(t_{c}^{-}) A_{2}(t_{c}^{-})/|v_{d1}|$,   
and the longitudinal factor is $c_{d2}^{(x)}W_{20}^{(x)}$ \cite{lp_form}. 
Application of our perturbative calculation to dimensions higher than 2 
shows that the form (\ref{da38}) remains valid for a general spatial dimension  
when the initial condition is separable in the longitudinal direction 
for both pulses. Equation (\ref{da37}) is also a generalization of the 
equation obtained for a fast two-pulse collision in the presence of 
weak quadratic loss in spatial dimension 1 [compare Eq. (\ref{da37}) 
with Eq. (32) in Ref. \cite{NHP2021}].  
We also observe that the longitudinal part in the expression for  
$\Delta A_{1}^{(c)}$, $c_{d2}^{(x)}W_{20}^{(x)}$, is universal 
in the sense that it does not depend on the details 
of the initial pulse shapes and on the collision time $t_{c}$. 
In contrast, the transverse part is not universal, since it does 
depend on the details of the initial pulse shapes and on $t_{c}$.  
Thus, the universality of the expression for the amplitude shift 
in the one-dimensional case, which was first demonstrated 
in Ref. \cite{NHP2021}, is extended to spatial dimension 2 
(and to spatial dimension $n$), but in a somewhat restricted manner. 
More specifically, in the two-dimensional (and the $n$-dimensional) case, 
only the overall and longitudinal parts of the expression for $\Delta A_{1}^{(c)}$ 
are universal, and this is true when the initial condition is separable in the 
longitudinal direction for both pulses.

\subsubsection{Evolution of $\phi_{1}(x,y,t)$ in the post-collision interval}
\label{dynamics2}

It is important to analyze the dynamics of $\phi_{1}(x,y,t)$ in the 
post-collision interval for a separable initial condition for the 
following reasons. First, for a separable initial condition, 
it is possible to prove that the pulse shape in the longitudinal 
direction is not changed at all by the collision and this requires 
analysis of $\phi_{1}(x,y,t)$ in the post-collision interval. 
Second, the simplest and clearest demonstration of the 
collision-induced change in pulse shape in the transverse 
direction is realized when the initial condition is separable 
for both pulses. Third, in both experiments and simulations 
of fast collisions, the change in the pulse shape can only 
be measured accurately in the post-collision interval.

The evolution of $\phi_{1}$ in the post-collision interval is 
described by the unperturbed linear diffusion equation (\ref{da21}).    
Using Eqs. (\ref{da22}) and (\ref{da33}), we find that the 
initial condition for Eq. (\ref{da21}) is    
\begin{eqnarray} &&
\phi_{1}(x,y,t_{c}^{+})=
-\tilde a_{1}(t_{c}^{-}) 
g_{1}^{(x)}(x,t_{c})g_{12}^{(y)}(y,t_{c}),
\label{da41}
\end{eqnarray}        
where 
\begin{eqnarray} &&
\tilde a_{1}(t_{c}^{-})=
2\epsilon_{2}A_{1}(t_{c}^{-}) A_{2}(t_{c}^{-}) 
c_{d2}^{(x)} W_{20}^{(x)}/|v_{d1}|, 
\label{da42}
\end{eqnarray}        
and 
\begin{eqnarray} &&
g_{12}^{(y)}(y,t_{c})= 
g_{1}^{(y)}(y,t_{c})g_{2}^{(y)}(y,t_{c}).
\label{da43}
\end{eqnarray}  
The Fourier transform of the initial condition (\ref{da41}) is 
\begin{eqnarray} &&
\hat \phi_{1}(k_{1},k_{2},t_{c}^{+})=
-\tilde a_{1}(t_{c}^{-}) \hat g_{1}^{(x)}(k_{1},t_{c}) 
\hat g_{12}^{(y)}(k_{2},t_{c}),
\label{da44}
\end{eqnarray}         
where $\hat g_{1}^{(x)}$ and $\hat g_{12}^{(y)}$ are the Fourier 
transforms of $g_{1}^{(x)}$ and $g_{12}^{(y)}$ with respect to 
$x$ and $y$, respectively. Substitution of Eq. (\ref{da44}) 
into Eq. (\ref{da23}) yields: 
\begin{eqnarray}&&
\phi_{1}(x,y,t)=
-\tilde a_{1}(t_{c}^{-}) 
{\cal F}^{-1}\left(\hat g_{1}^{(x)}(k_{1},t_{c})
\exp[- k_{1}^{2}(t-t_{c})]\right)
\nonumber \\&&
\times {\cal F}^{-1}\left(\hat g_{12}^{(y)}(k_{2},t_{c})
\exp[- k_{2}^{2}(t-t_{c})]\right). 
\label{da45}
\end{eqnarray} 
When the initial condition for pulse 1 is separable, 
$\hat g_{1}^{(x)}(k_{1},t_{c})\exp[- k_{1}^{2}(t-t_{c})]=\hat g_{1}^{(x)}(k_{1},t)$. 
Using this relation in Eq. (\ref{da45}), we obtain the following 
expression for $\phi_{1}(x,y,t)$ in the post-collision interval 
for a separable initial condition:  
\begin{eqnarray}&&
\phi_{1}(x,y,t)=
-\tilde a_{1}(t_{c}^{-}) g_{1}^{(x)}(x,t)
{\cal F}^{-1}\left(\hat g_{12}^{(y)}(k_{2},t_{c})
\exp[- k_{2}^{2}(t-t_{c})]\right). 
\label{da46}
\end{eqnarray} 
Thus, when the initial condition is separable for both pulses, 
the $x$ dependences of $\phi_{1}(x,y,t)$ and $\tilde u_{10}(x,y,t)$ 
are identical for $t > t_{c}$. It follows that the pulse shape 
in the longitudinal direction is not changed at all by the collision. 
Furthermore, we see that the change in the pulse shape in the 
transverse direction in the post-collision interval is 
proportional to the inverse Fourier transform of 
$\hat g_{12}^{(y)}(k_{2},t_{c})\exp[- k_{2}^{2}(t-t_{c})]$.

\section{Demonstration of new collision-induced effects in spatial dimension 2}
\label{simu}

\subsection{Introduction}
\label{simu_intro}

Let us use the perturbation method of section \ref{da_2D}
together with numerical simulations with Eq. (\ref{da1}) to demonstrate 
the following three major effects and properties of the collision, which exist 
only in spatial dimension higher than 1. (1) The universality of the 
longitudinal part in the expression for the collision-induced amplitude shift, 
an attribute that can be useful in the design of multisequence chemical 
communication links. (2) The effect of anisotropy in the initial condition. 
(3) The collision-induced change in the pulse shape in the transverse direction. 
For each collisional effect, we first use our perturbation method to obtain 
explicit approximate formulas that demonstrate the effect.  
We then carry out extensive numerical simulations with 
Eq. (\ref{da1}) and compare the simulations results with the 
approximate predictions of the perturbation method for each 
of the three collisional effects. 
We solve Eq. (\ref{da1}) numerically by the split-step method 
with periodic boundary conditions \cite{Verwer2003}.

\subsection{Universality of the longitudinal part in the 
expression for the amplitude shift} 
\label{simu_universality}

In subsection \ref{da_separable_IC}, we showed that for an initial condition 
that is separable for both pulses, the longitudinal part in the expression 
for the collision-induced amplitude shift is universal in the sense that it 
does not depend on the details of the initial pulse shapes. 
In the current subsection, we demonstrate this property. 
For this purpose, we first obtain explicit formulas for $\Delta A_{1}^{(c)}$ 
for two initial pulse shapes that have very different dependences on 
the $x$ coordinate. We then verify the validity of the formulas for 
$\Delta A_{1}^{(c)}$ by extensive numerical simulations with 
Eq. (\ref{da1}). These simulations are very important, 
since they show that the approximations used in our perturbative 
calculation are indeed valid for very different pulse shapes.
In this way, the simulations validate the theoretical prediction 
for universality of the expression for the amplitude shift.

The initial $x$ dependence for the first pulse type is Gaussian, 
i.e., it is rapidly decreasing with increasing value of $|x-x_{j0}|$.                      
In contrast, the initial $x$ dependence for the second pulse type 
is given by a Cauchy-Lorentz distribution, i.e., it decreases 
slowly (as a power-law) with increasing value of $|x-x_{j0}|$.  
The initial pulse profile in the transverse direction is taken as Gaussian, 
since this choice enables the explicit calculation of the integral with respect 
to $y$ on the right hand side of Eq. (\ref{da37}). Therefore, 
the two initial conditions that we consider are:  
\begin{eqnarray}&&
u_{1}(x,y,0)=A_{1}(0)\exp \left[-\frac{x^{2}}{2W^{(x)2}_{10}}
-\frac{y^{2}}{2W^{(y)2}_{10}} \right],
\nonumber \\&&
u_{2}(x,y,0)=A_{2}(0)\exp \left[-\frac{(x-x_{20})^{2}}{2W^{(x)2}_{20}}
-\frac{y^{2}}{2W^{(y)2}_{20}} \right], 
\label{da51}
\end{eqnarray}  
for Gaussian pulses, and      
 \begin{eqnarray}&&
u_{1}(x,y,0)=A_{1}(0)\left[1 + \frac{2x^{4}}{W^{(x)4}_{10}} \right]^{-1}
\exp \left[-\frac{y^{2}}{2W^{(y)2}_{10}} \right],
\nonumber \\&&
u_{2}(x,y,0)=A_{2}(0)\left[1 + \frac{2(x-x_{20})^{4}}{W^{(x)4}_{20}} \right]^{-1}
\exp \left[-\frac{y^{2}}{2W^{(y)2}_{20}} \right],
\label{da52}
\end{eqnarray}    
for Cauchy-Lorentz-Gaussian pulses.

We first obtain the expression for the collision-induced amplitude shift 
for the initial conditions (\ref{da51}) and (\ref{da52}). From 
Eq. (\ref{appC_5}) it follows that for both initial conditions 
\begin{eqnarray}&&
g_{j}^{(y)}(y,t_{c})=
\frac{W_{j0}^{(y)}}
{(W_{j0}^{(y)2} + 2t_{c})^{1/2}}
\exp\left[-\frac{y^{2}}{2W_{j0}^{(y)2} + 4t_{c}} \right]. 
\label{da53}
\end{eqnarray}   
In addition, $c_{d1}^{(y)}=(2\pi)^{1/2}$ in both cases. 
We substitute Eq. (\ref{da53}) and the value of $c_{d1}^{(y)}$ 
into Eq. (\ref{da37}), and carry out the integration with 
respect to $y$. We obtain the following expression for 
$\Delta A_{1}^{(c)}$: 
\begin{eqnarray} &&
\!\!\!\!
\Delta A_{1}^{(c)}=-\frac{2\epsilon_{2}
A_{1}(t_{c}^{-}) A_{2}(t_{c}^{-})}{|v_{d1}|}
\frac{c_{d2}^{(x)}W_{20}^{(x)}W_{20}^{(y)}}
{(W_{10}^{(y)2} + W_{20}^{(y)2} + 4t_{c})^{1/2}},
\label{da54}     
\end{eqnarray} 
where $c_{d2}^{(x)}=(2\pi)^{1/2}$ for Gaussian pulses, and 
$c_{d2}^{(x)}=\pi/2^{3/4}$ for Cauchy-Lorentz-Gaussian pulses. 
We observe that the longitudinal part in the expression 
for $\Delta A_{1}^{(c)}$, $c_{d2}^{(x)}W_{20}^{(x)}$, is universal. 
In contrast, the transverse part, which is given by: 
\begin{eqnarray} &&
\!\!\!\!\!\!\!\!\!
\mbox{transverse factor}=
\frac{W_{20}^{(y)}}
{(W_{10}^{(y)2} + W_{20}^{(y)2} + 4t_{c})^{1/2}},
\label{da55}
\end{eqnarray}                                   
depends on $t_{c}$, and does not possess a simple universal form 
like the longitudinal part. An important aspect of the nonuniversal 
nature of the equation for $\Delta A_{1}^{(c)}$ in spatial dimension 2 
is the deviation of the dependence on $|v_{d1}|$ from the $1/|v_{d1}|$ 
scaling that exists in the one-dimensional case \cite{PNH2017B,NHP2021}, 
and also in fast collisions between solitons of the nonlinear 
Schr\"odinger equation in the presence of weak 
nonlinear dissipation in spatial dimension 1 \cite{PNC2010,PC2012}. 
Indeed, the collision time $t_{c}$ is given by $t_{c}=(x_{10}-x_{20})/v_{d1}$.  
Therefore, the deviation of the $|v_{d1}|$ dependence of $\Delta A_{1}^{(c)}$ 
from the $1/|v_{d1}|$ scaling is due to the term $4t_{c}=4(x_{10}-x_{20})/v_{d1}$ 
in the factor $(W_{10}^{(y)2} + W_{20}^{(y)2} + 4(x_{10}-x_{20})/v_{d1})^{1/2}$    
on the right hand side of Eq. (\ref{da54}). To characterize the latter deviation, 
we define the quantity $\Delta A_{1}^{(c)(s)}$, which is the approximate expression 
for the amplitude shift that is obtained from the full expression by neglecting the 
$4(x_{10}-x_{20})/v_{d1}$ term. By this definition, $\Delta A_{1}^{(c)(s)}$ 
is given by: 
\begin{eqnarray} &&
\!\!\!\!
\Delta A_{1}^{(c)(s)}=-\frac{2\epsilon_{2}
A_{1}(t_{c}^{-}) A_{2}(t_{c}^{-})}{|v_{d1}|}
\frac{c_{d2}^{(x)}W_{20}^{(x)}W_{20}^{(y)}}
{(W_{10}^{(y)2} + W_{20}^{(y)2})^{1/2}}.
\label{da56}     
\end{eqnarray} 
Thus, the difference $|\Delta A_{1}^{(c)} - \Delta A_{1}^{(c)(s)}|$
is a measure for the departure of the $|v_{d1}|$ dependence of 
$\Delta A_{1}^{(c)}$ from the $1/|v_{d1}|$ scaling observed in 
the one-dimensional case. Since in a complete collision $|x_{20}-x_{10}| \gg 1$,  
the term $4(x_{10}-x_{20})/v_{d1}$ is not necessarily small for intermediate 
values of $|v_{d1}|$. As a result, the departure from the $1/|v_{d1}|$ scaling 
might be substantial even for intermediate $|v_{d1}|$ values.

We check the perturbation theory predictions for universality of the longitudinal 
part in the expression for $\Delta A_{1}^{(c)}$ by performing numerical simulations 
with Eq. (\ref{da1}) with the two initial conditions (\ref{da51}) and (\ref{da52}). 
The extensive simulations with these initial conditions provide a careful test for the 
validity of the perturbation theory approximations for widely different pulse shapes.  
In this manner, the simulations help validate the universality of the longitudinal part 
in the expression for $\Delta A_{1}^{(c)}$. 
We are interested in fast collisions in the presence of weak quadratic loss. 
Therefore, we carry out the simulations with $\epsilon_{2}=0.01$ and with 
$v_{d1}$ values in the intervals $4 \le |v_{d1}| \le 60$. 
The parameter values of the initial conditions (\ref{da51}) 
and (\ref{da52}) are $A_{j}(0)=1$, $x_{20}=\pm 20$, $W_{10}^{(x)}=3$, 
$W_{10}^{(y)}=5$, $W_{20}^{(x)}=4$, and $W_{20}^{(y)}=6$.  
The final time is $t_{f}=2t_{c}=-2x_{20}/v_{d1}$. The values of $x_{20}$ and $t_{f}$ 
ensure that the colliding pulses are well separated at $t=0$ and at $t=t_{f}$. 
We emphasize that results similar to the ones described 
here are obtained in simulations with other parameter values. 
For each initial condition, we compare the dependence of $\Delta A_{1}^{(c)}$ 
on $v_{d1}$ obtained in the simulations with the perturbation theory prediction 
of Eq. (\ref{da54}), and with the crude approximation of Eq. (\ref{da56}). 
We also discuss the behavior of the relative errors in the approximation 
of the amplitude shift (in percentage), which are defined 
by $E_{r}^{(1)}=|\Delta A_{1}^{(c)(num)}-\Delta A_{1}^{(c)(th)}|
\times 100/|\Delta A_{1}^{(c)(th)}|$ and 
$E_{r}^{(2)}=|\Delta A_{1}^{(c)(num)}-\Delta A_{1}^{(c)(s)}|
\times 100/|\Delta A_{1}^{(c)(s)}|$, respectively.

\begin{figure}[ptb]
\begin{center}
\begin{tabular}{cc}
\epsfxsize=9.0cm  \epsffile{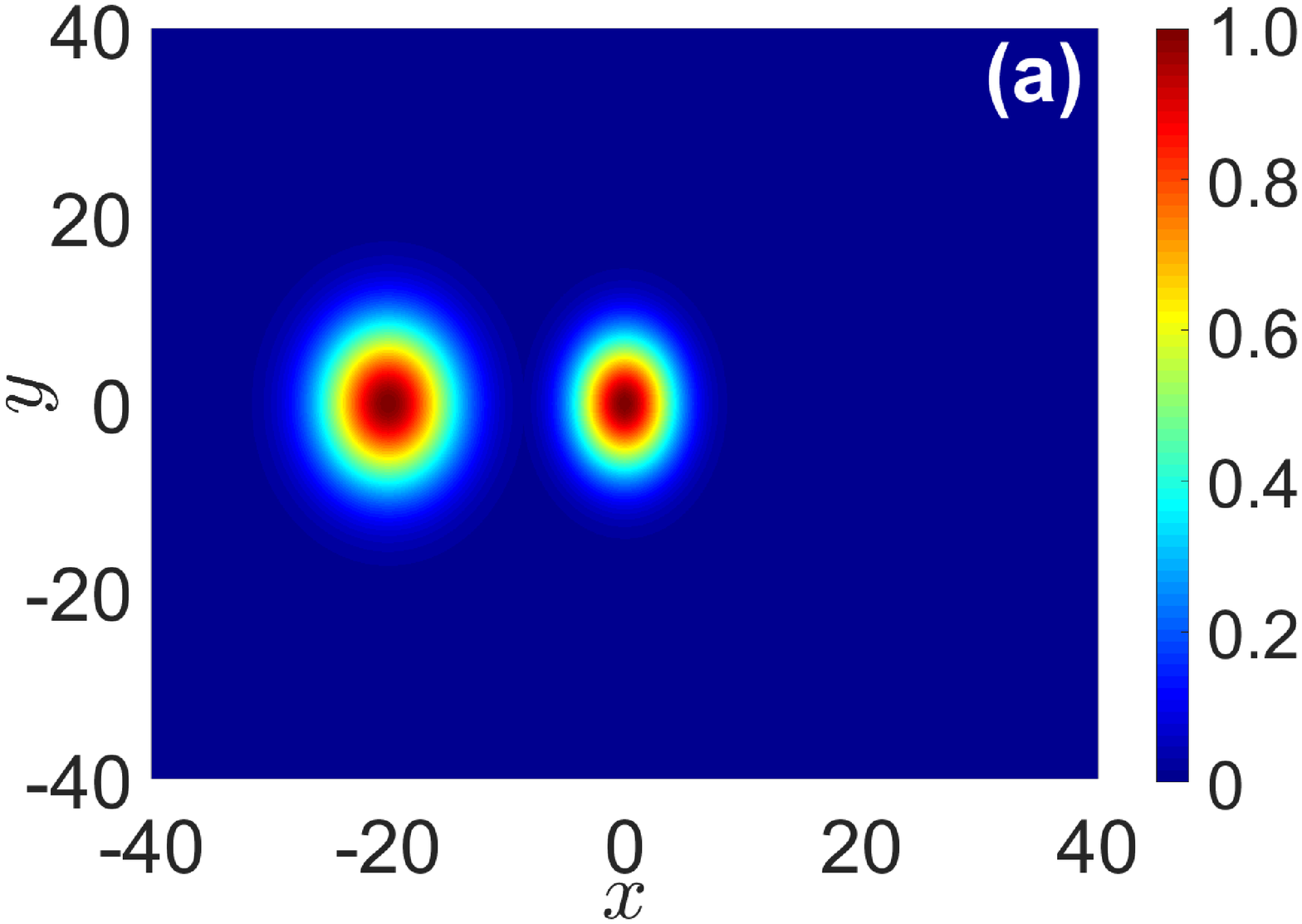} \\
\epsfxsize=9.0cm  \epsffile{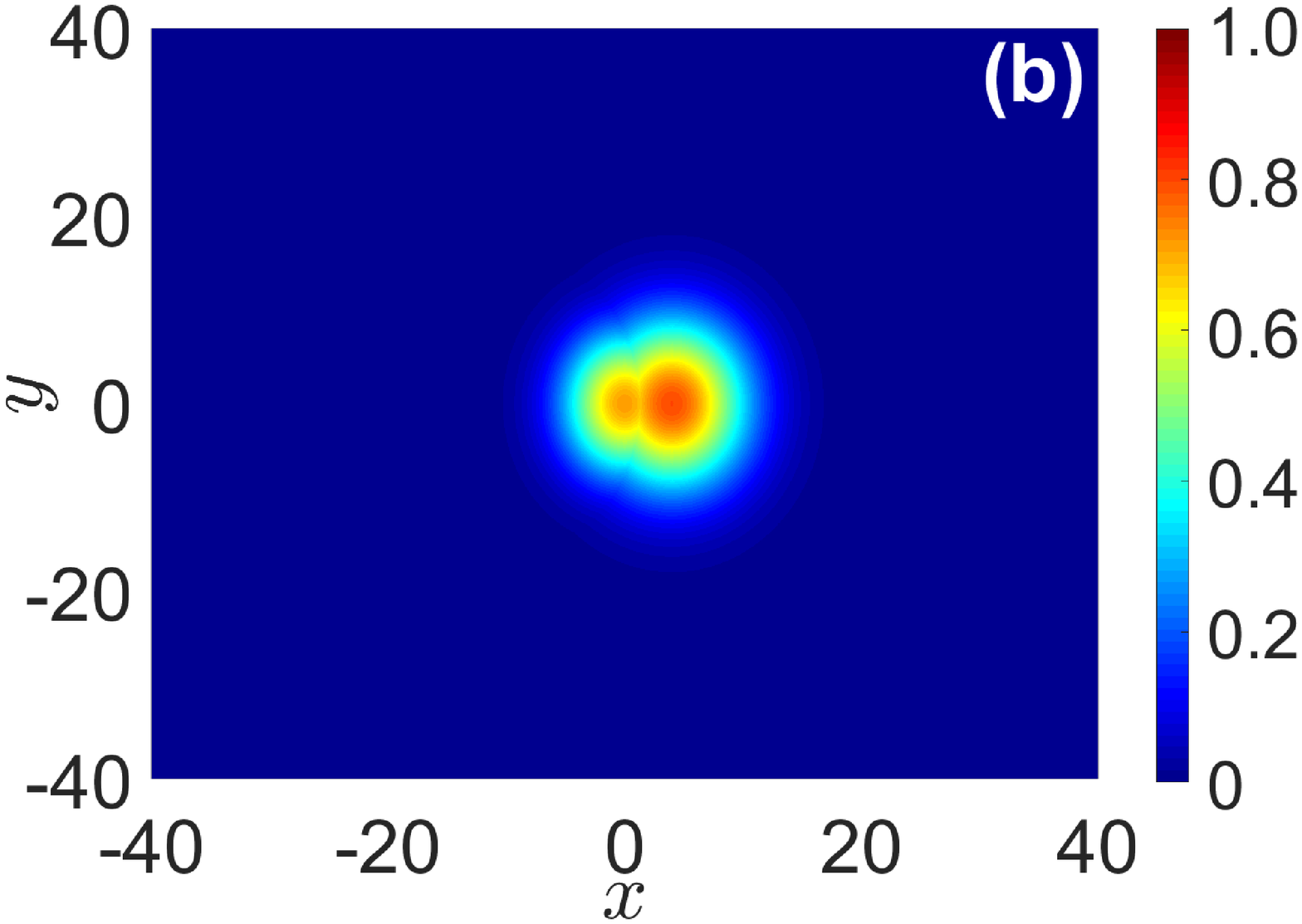} \\
\epsfxsize=9.0cm  \epsffile{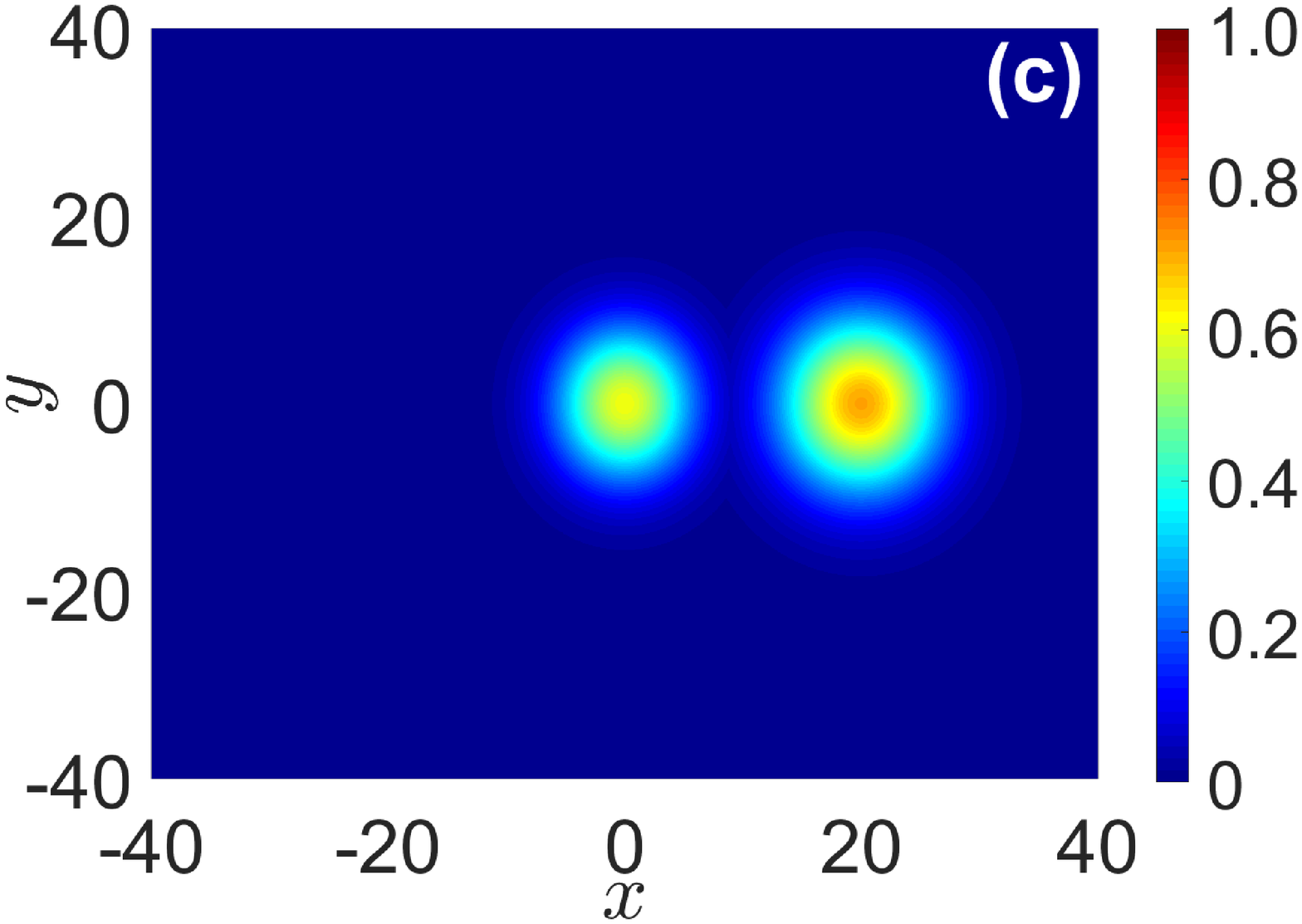} 
\end{tabular}
\end{center}
\caption{(Color online) 
Contour plots of the  pulse shapes $u_{j}(x,y,t)$ at $t=0$ (a), 
$t=t_{i}=2.4$ (b), and $t=t_{f}=4$ (c) in a fast collision between 
two Gaussian pulses with parameter values $\epsilon_{2}=0.01$ and $v_{d1}=10$.
The plots represent the pulse shapes obtained by numerical solution 
of Eq. (\ref{da1}) with the initial condition (\ref{da51}).}          
\label{fig1}
\end{figure}

\begin{figure}[ptb]
\begin{center}
\epsfxsize=11cm \epsffile{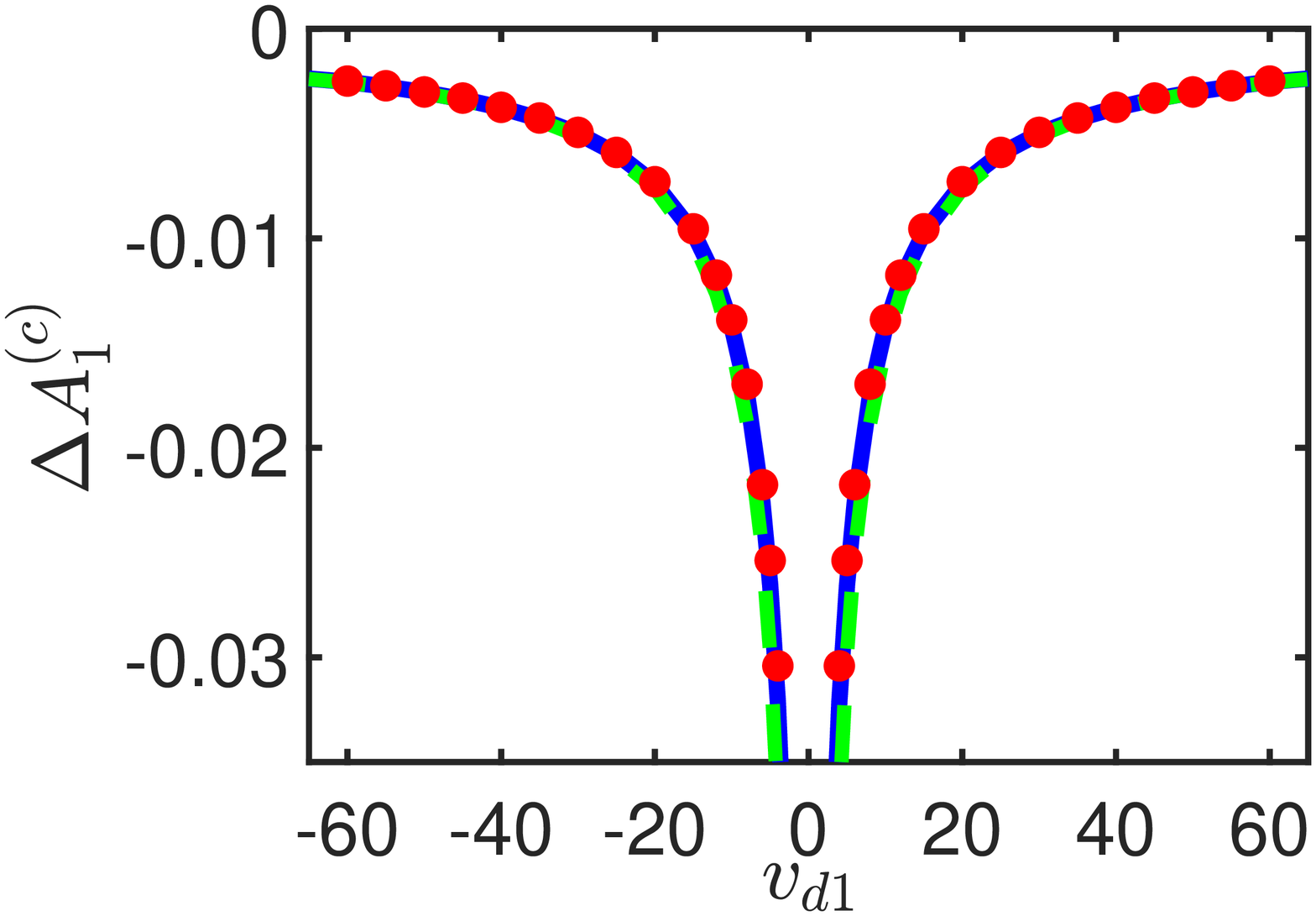}
\end{center}
\caption{(Color online) 
The collision-induced amplitude shift of pulse 1 
$\Delta A_{1}^{(c)}$ vs advection velocity $v_{d1}$ 
in a fast collision between two Gaussian pulses for 
$\epsilon_{2}=0.01$. The red circles represent 
the result obtained by numerical simulations with Eq. (\ref{da1}) 
with the initial condition (\ref{da51}). 
The solid blue and dashed green curves correspond to the theoretical   
predictions of Eqs. (\ref{da54}) and (\ref{da56}), respectively.}
\label{fig2}
\end{figure}

We first discuss the results of the simulations for fast collisions between 
pulses with rapidly decaying tails, which are represented by Gaussian 
pulses. The initial pulse shapes $u_{j}(x,y,0)$, and the pulse  
shapes $u_{j}(x,y,t)$ obtained in the simulation with $v_{d1}=10$
at the intermediate time $t_{i}=2.4>t_{c}$ \cite{ti_values}, and at 
the final time $t_{f}=4$ are shown in Fig. \ref{fig1}. 
We observe that the pulses experience broadening due to diffusion. 
We also observe that the maximum values of $u_{j}(x,y,t)$  
decrease with increasing time, mainly due to diffusion. 
The dependence of $\Delta A_{1}^{(c)}$ on $v_{d1}$ obtained 
in the simulations is shown in Fig. \ref{fig2} together with the analytic 
predictions $\Delta A_{1}^{(c)}$ and $\Delta A_{1}^{(c)(s)}$ of Eqs. (\ref{da54}) 
and (\ref{da56}). The agreement between the simulations result and 
the analytic prediction of Eq. (\ref{da54}) is very good 
despite the diffusion-induced pulse broadening.    
More specifically, the relative error $E_{r}^{(1)}$ is 
smaller than $2.4\%$ for $10 \le |v_{d1}| \le 60$ and 
smaller than $5.4\%$ for $4 \le |v_{d1}| <10$. 
We also note that the relative error $E_{r}^{(2)}$ is  
smaller than $8.2\%$ for $10 \le |v_{d1}| \le 60$ 
and smaller than $17.9\%$ for $4 \le |v_{d1}| <10$. 
The latter values are noticeably larger than the 
corresponding values of $E_{r}^{(1)}$. Thus, in accordance 
with the perturbation theory prediction, the deviation of the 
$v_{d1}$ dependence of $\Delta A_{1}^{(c)}$ from the $1/|v_{d1}|$ 
scaling is noticeable already at intermediate $|v_{d1}|$ values, 
and it increases with decreasing values of $|v_{d1}|$.

\begin{figure}[ptb]
\begin{center}
\begin{tabular}{cc}
\epsfxsize=9.0cm  \epsffile{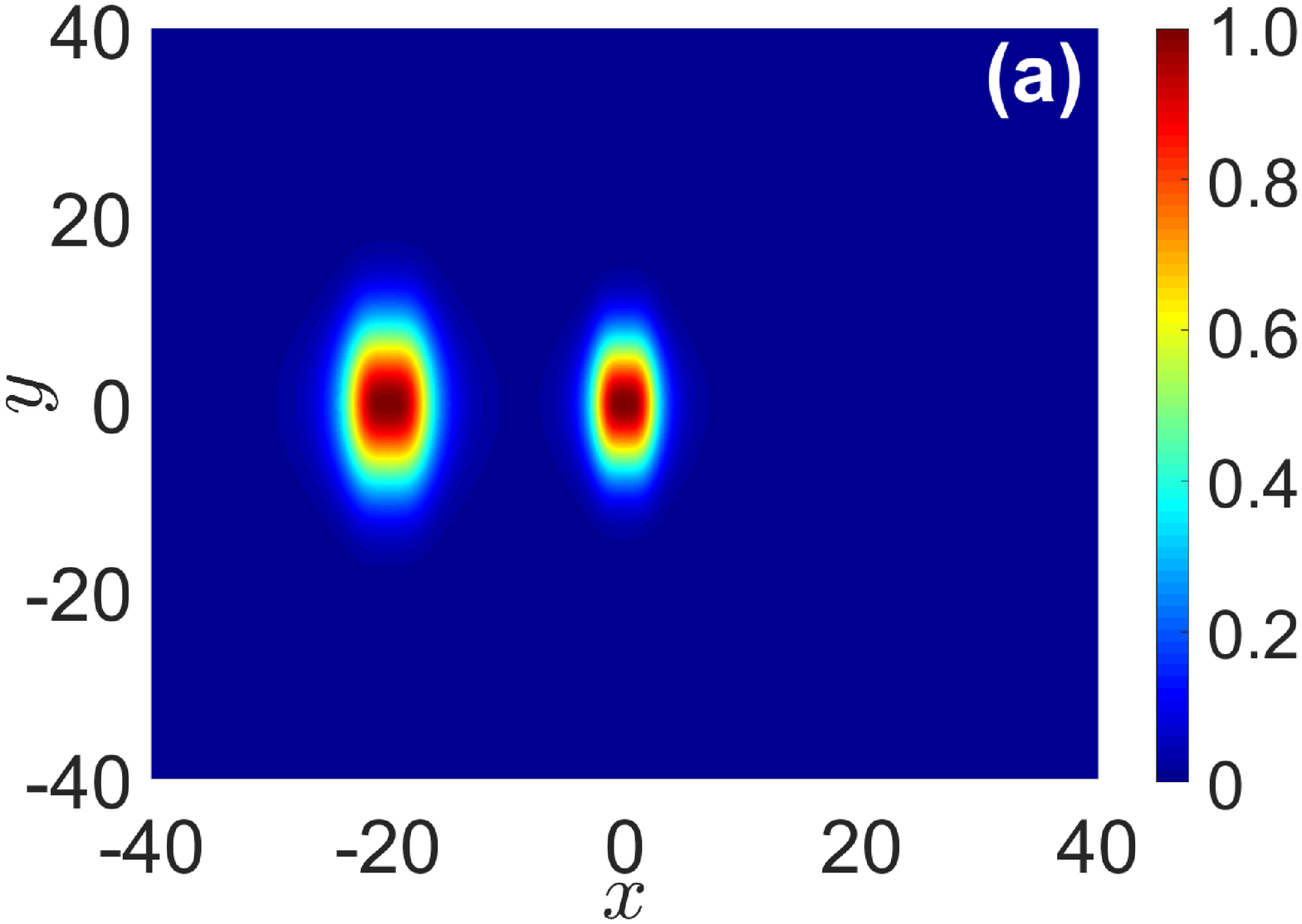} \\
\epsfxsize=9.0cm  \epsffile{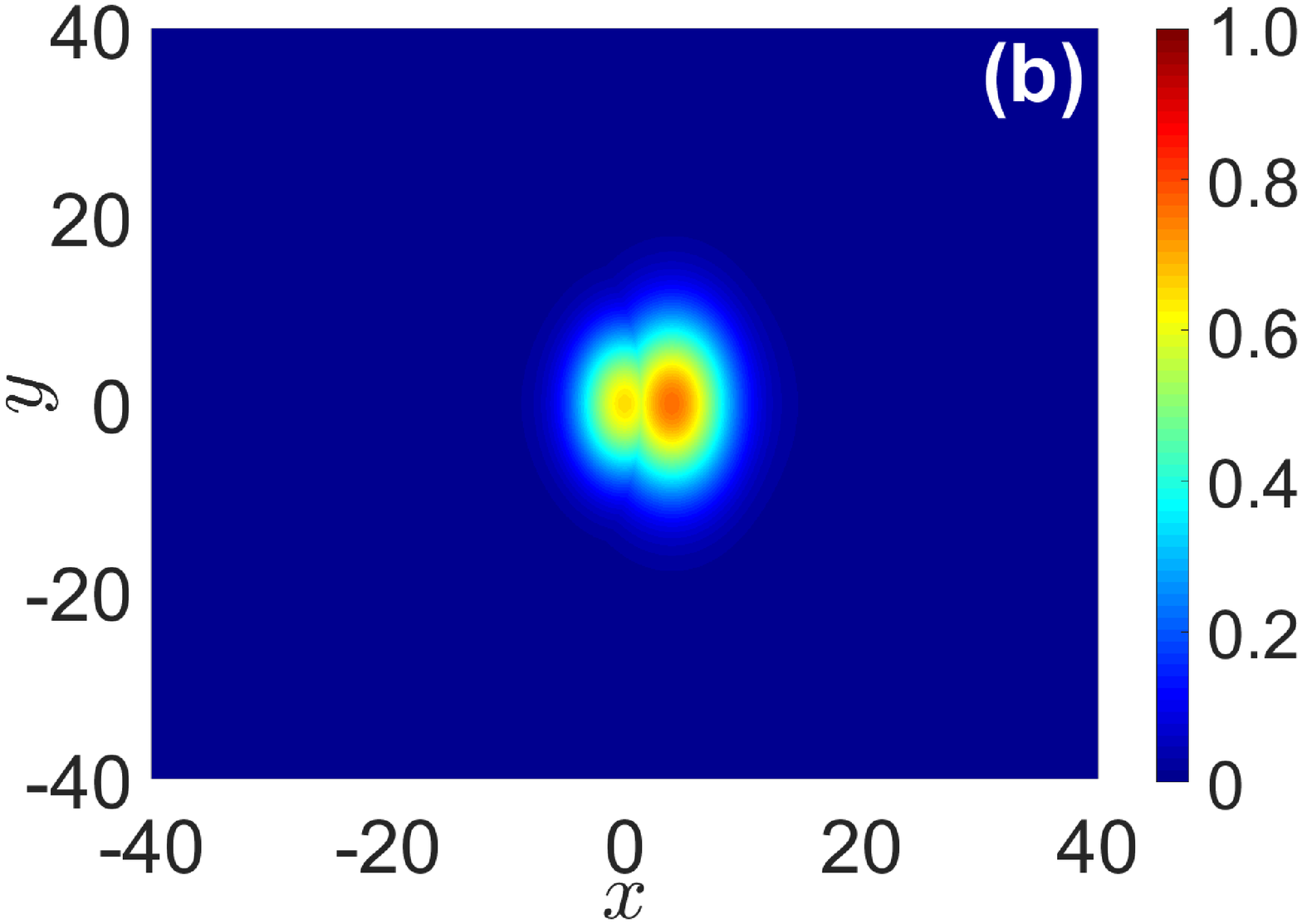} \\
\epsfxsize=9.0cm  \epsffile{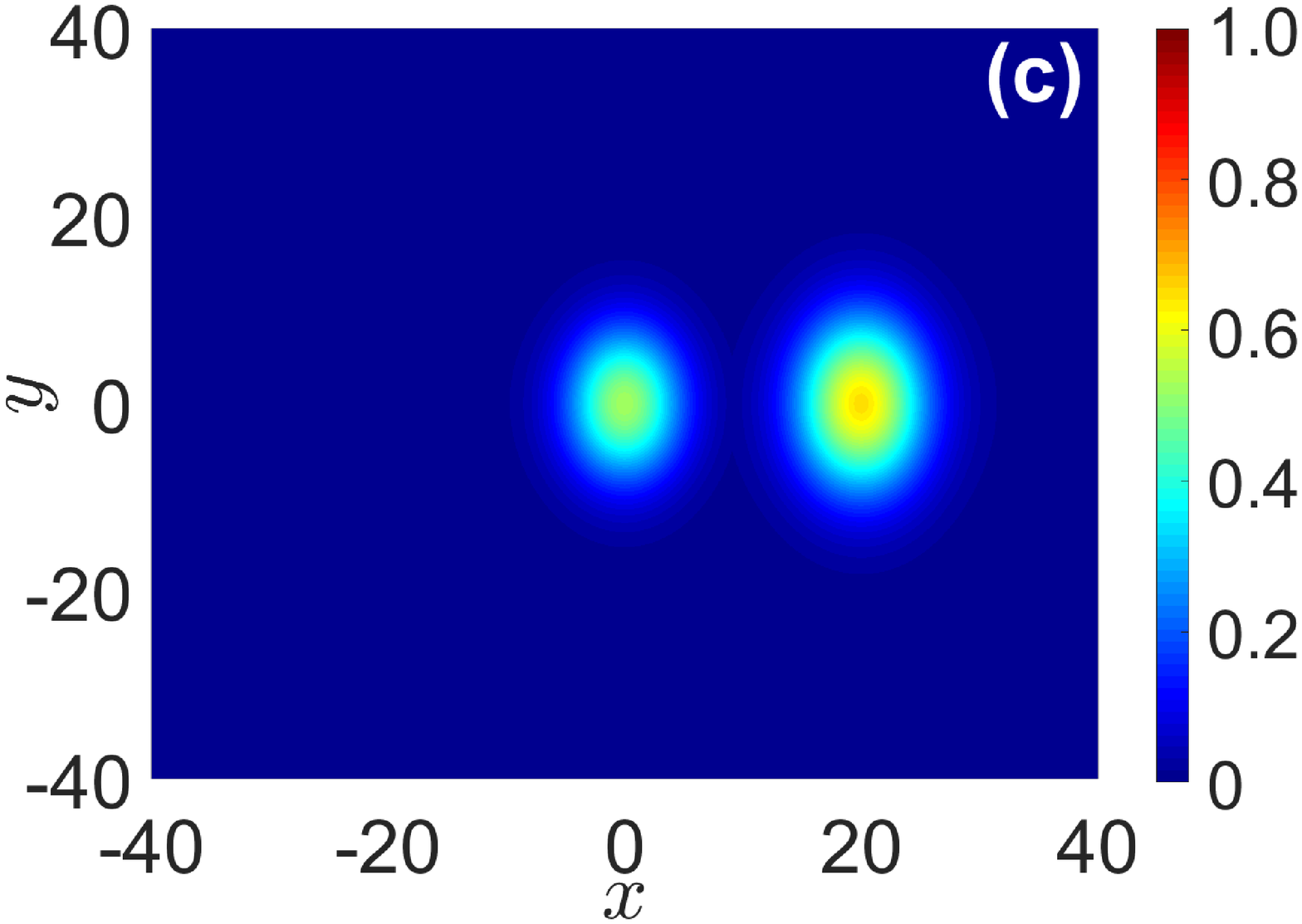} 
\end{tabular}
\end{center}
\caption{(Color online) 
Contour plots of the  pulse shapes $u_{j}(x,y,t)$ at $t=0$ (a), 
$t=t_{i}=2.4$ (b), and $t=t_{f}=4$ (c) in a fast collision between 
two Cauchy-Lorentz-Gaussian pulses with parameter values 
$\epsilon_{2}=0.01$ and $v_{d1}=10$.
The plots represent the pulse shapes obtained by numerical solution 
of Eq. (\ref{da1}) with the initial condition (\ref{da52}).}          
\label{fig3}
\end{figure}

\begin{figure}[ptb]
\begin{center}
\epsfxsize=11cm \epsffile{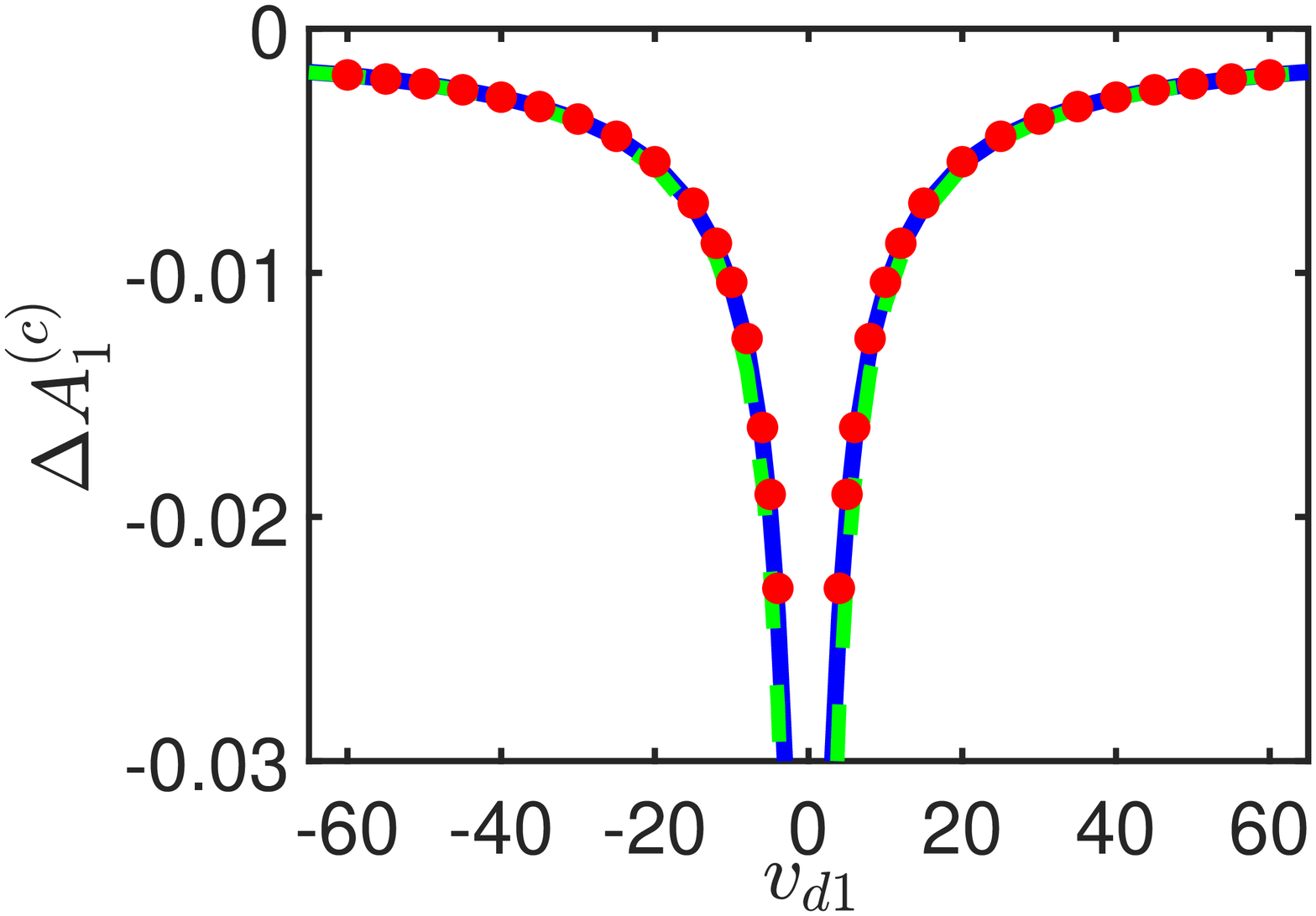}
\end{center}
\caption{(Color online) 
The collision-induced amplitude shift of pulse 1 
$\Delta A_{1}^{(c)}$ vs advection velocity $v_{d1}$ 
in a fast collision between two Cauchy-Lorentz-Gaussian
pulses for $\epsilon_{2}=0.01$. The red circles represent 
the result obtained by numerical simulations with Eq. (\ref{da1}) 
with the initial condition (\ref{da52}). 
The solid blue and dashed green curves represent the theoretical   
predictions of Eqs. (\ref{da54}) and (\ref{da56}), respectively.}
\label{fig4}
\end{figure}

Let us discuss the simulations results for fast collisions 
between pulses with slowly decaying tails in the longitudinal direction, 
which are represented by Cauchy-Lorentz-Gaussian pulses. 
In this case, it is unclear if the sharp-peak approximation  
that is used in the derivation of Eq. (\ref{da14}) is valid.    
Therefore, the simulations of fast collisions between 
Cauchy-Lorentz-Gaussian pulses serve as an important check 
of both the perturbation theory approximations and the perturbation 
theory prediction for universality of the longitudinal part in 
the expression for $\Delta A_{1}^{(c)}$.    
Figure \ref{fig3} shows the pulse shapes $u_j(x,y,t)$ obtained 
in the simulation with $v_{d1}=10$ at $t=0$, $t_{i}=2.4$, and $t_{f}=4$. 
It is seen that the pulses undergo significant broadening due 
to diffusion, and that the maximum values of $u_{j}(x,y,t)$  
decrease with increasing $t$ as a result. The numerically 
obtained dependence of $\Delta A_{1}^{(c)}$ on $v_{d1}$ 
is shown in Fig. \ref{fig4} along with the perturbation theory 
predictions of Eqs. (\ref{da54}) and (\ref{da56}). 
We observe very good agreement between the simulations result and 
the analytic prediction of Eq. (\ref{da54}) despite the significant 
pulse broadening. In particular, the relative error $E_{r}^{(1)}$ is 
smaller than $2.1\%$ for $10 \le |v_{d1}| \le 60$ and 
smaller than $4.5\%$ for $4 \le |v_{d1}| <10$.   
These values are comparable to the values of $E_{r}^{(1)}$ 
for collisions between Gaussian pulses. Thus, based on the results 
shown in Figs. \ref{fig2} and \ref{fig4} and on similar results 
obtained with other values of the physical parameters, 
we conclude that the longitudinal part in the expression for 
$\Delta A_{1}^{(c)}$ is indeed universal in the sense that it is 
not very sensitive to the details of the initial pulse shapes. 
We also note that the values of $E_{r}^{(2)}$ obtained in collisions 
between Cauchy-Lorentz-Gaussian pulses are noticeably larger than the 
corresponding values of $E_{r}^{(1)}$ for intermediate and small 
$|v_{d1}|$ values. Therefore, our simulations also demonstrate 
that the departure of the $v_{d1}$ dependence of $\Delta A_{1}^{(c)}$ 
from the $1/|v_{d1}|$ scaling is noticeable already at 
intermediate values of $|v_{d1}|$.

\subsection{Dependence of the amplitude shift on the inter-pulse orientation angle} 
\label{simu_angle}

Another phenomenon that exists only in spatial dimension 
higher than 1 is associated with the effects of anisotropy 
in the physical system. In the current paper, we are interested 
in studying the effects of anisotropy in the initial condition. 
In a simple setup, this anisotropy can be characterized by a 
single angle $\theta_{0}$, e.g., the angle between a 
``preferred'' direction in the initial condition and the $x$ axis.  
Consider as an example the case where the initial width of pulse 1 
along one direction, which we denote by $x'$, is larger than the 
initial width along the perpendicular direction $y'$. We then define $\theta_{0}$  
as the angle that the $x'$ axis forms with the $x$ axis of 
our coordinate system. Therefore, $\theta_{0}$ is the angle 
between the advection velocity vector and the $x'$ axis. 
In addition, if pulse 2 is circularly symmetric, or is elongated 
along the $x$ or the $y$ axes, then $\theta_{0}$ can also be 
regarded as the orientation angle between the two pulses. 
A central question about the collision dynamics in this 
setup concerns the dependence of the amplitude shift 
$\Delta A_{1}^{(c)}$ on the orientation angle $\theta_{0}$. 
In the current subsection, we address this central question  
both analytically and by numerical simulations.

\begin{figure}[ptb]
\begin{center}
\epsfxsize=9cm  \epsffile{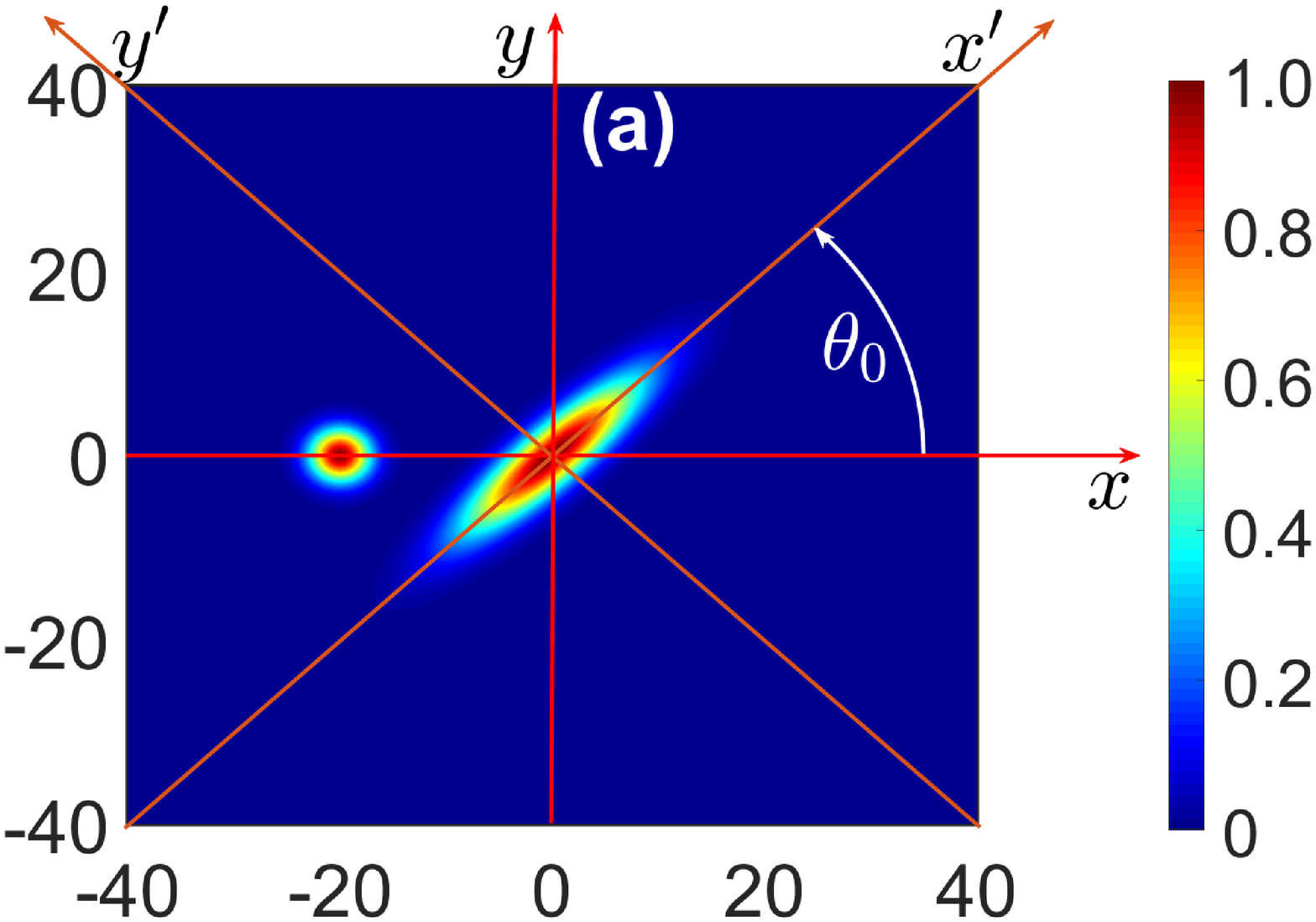} \\
\epsfxsize=9cm  \epsffile{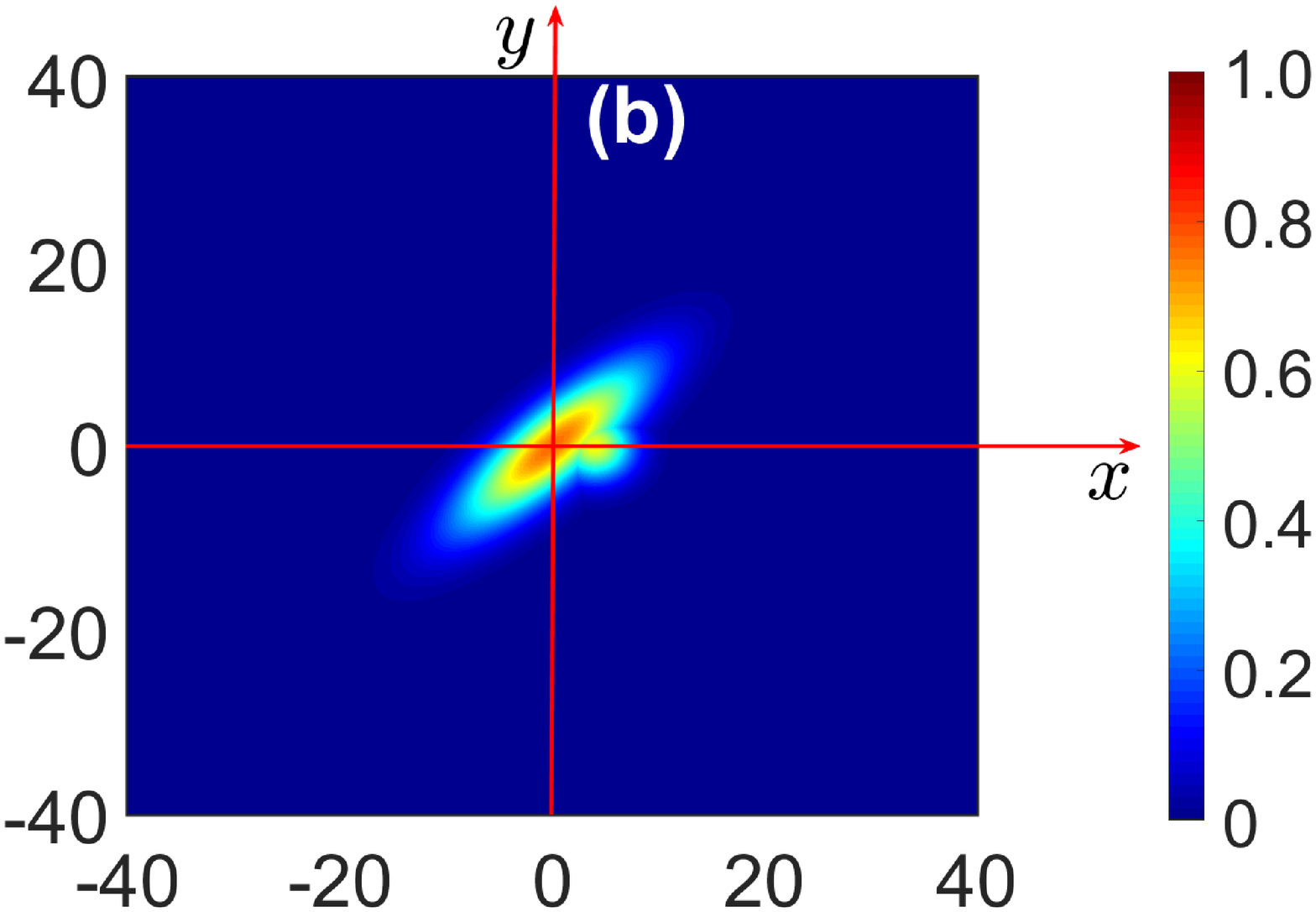} \\
\epsfxsize=9cm  \epsffile{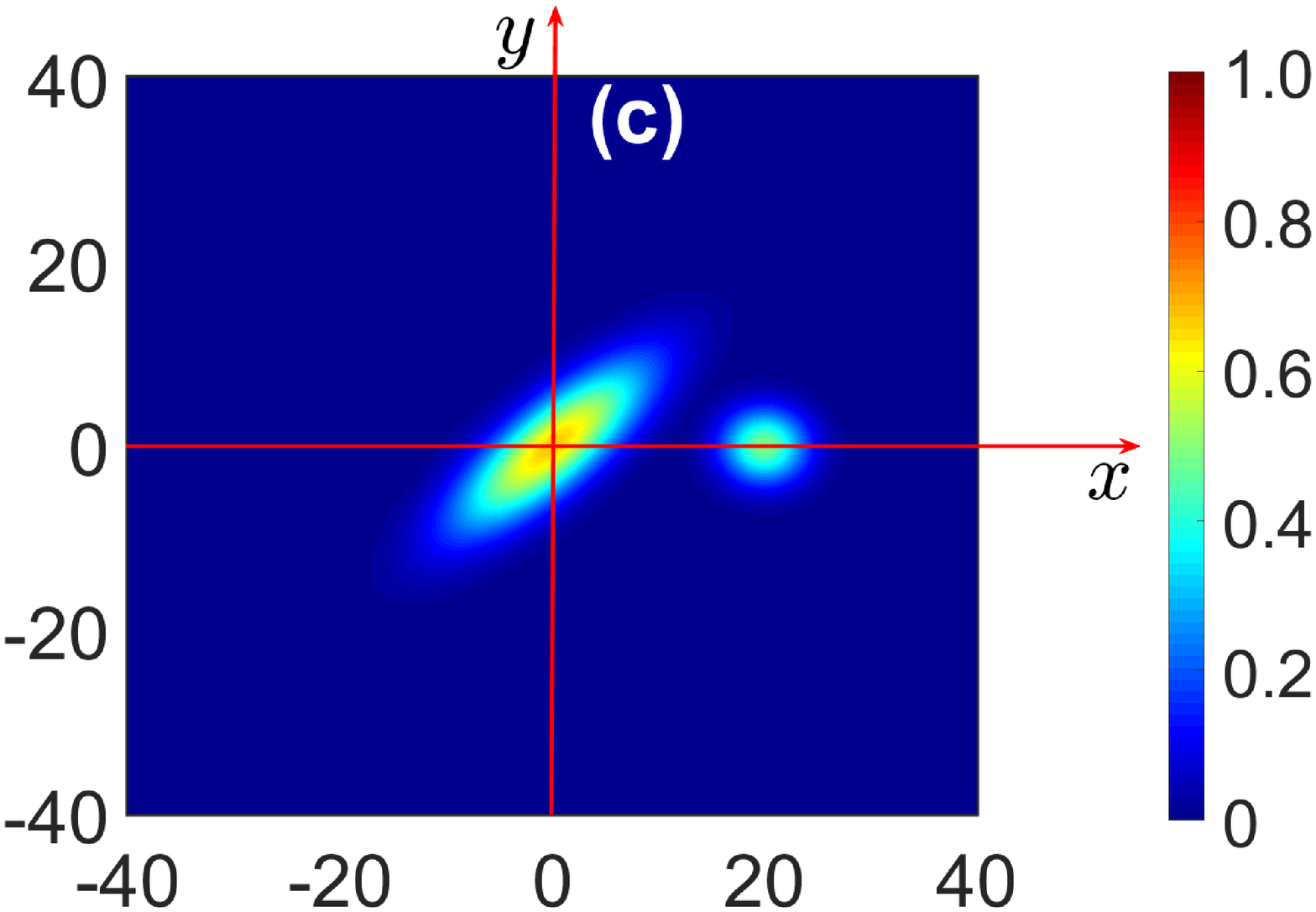} 
\end{center}
\caption{(Color online) 
Contour plots of the pulse shapes $u_{j}(x,y,t)$ at $t=0$ (a), 
$t=t_{i}=1.2$ (b), and $t=t_{f}=2$ (c) in a fast collision 
between two Gaussian pulses with the anisotropic initial 
condition of Eqs. (\ref{da62}) and (\ref{da64}). 
The orientation angle is $\theta_{0}=\pi/4$. 
The plots in (b) and (c) represent the pulse shapes 
obtained by numerical solution of Eq. (\ref{da1}) with 
physical parameter values $\epsilon_{2}=0.01$ and $v_{d1}=20$.}
\label{fig5}
\end{figure}

We consider an anisotropic collision setup, consisting  
of two initially well separated Gaussian pulses. In addition, 
pulse 1 is elongated along its $x'$ axis, which forms an angle $\theta_{0}$ 
with the $x$ axis, while pulse 2 is circularly symmetric. 
Figure \ref{fig5}(a) shows the contour plot of $u_{j}(x,y,0)$ 
for this anisotropic initial condition in the case $\theta_{0}=\pi/4$. 
The initial condition can be expressed as: 
\begin{eqnarray}&&
\!\!\!\!\!\!
u_{1}'(x',y',0)=A_{1}(0)\exp \left[ -\frac{x^{\prime 2}}{2W^{(x)2}_{10}}
-\frac{y^{\prime 2}}{2W^{(y)2}_{10}} \right],
\label{da61}
\end{eqnarray}  
and 
\begin{eqnarray}&&
\!\!\!\!\!\!
u_{2}(x,y,0)=A_{2}(0)\exp \left[ -\frac{(x-x_{20})^{2}}{2W^{2}_{20}}
-\frac{y^{2}}{2W^{2}_{20}} \right], 
\label{da62}
\end{eqnarray}   
where $u_{1}'(x',y',t)$ is the concentration of pulse 1 in the 
$(x',y',t)$ coordinate system, $W_{10}^{(x)} > W_{10}^{(y)}$, and 
\begin{eqnarray}&&
x' = x\cos\theta_{0} + y\sin\theta_{0},
\nonumber \\&&
y' = -x\sin\theta_{0} + y\cos\theta_{0}. 
\label{da63}
\end{eqnarray}  
Substituting Eq. (\ref{da63}) into Eq. (\ref{da61}), we obtain: 
\begin{eqnarray}&&
u_{1}(x,y,0)=A_{1}(0)\exp\left[ -B_{1}x^2 - B_{2}y^2 - B_{3}xy \right],
\label{da64}
\end{eqnarray}     
where 
\begin{eqnarray}&&
B_{1}=\frac{\cos^2 \theta_0}{2W_{10}^{(x)2}} + 
\frac{\sin^2 \theta_0}{2W_{10}^{(y)2}}, 
\nonumber
\end{eqnarray}     
\begin{eqnarray}&&
B_{2}=\frac{\sin^2 \theta_0}{2W_{10}^{(x)2}} + 
\frac{\cos^2 \theta_0}{2W_{10}^{(y)2}}, 
\nonumber
\end{eqnarray}     
and  
\begin{eqnarray}&&
B_{3}=\left(\frac{1}{W_{10}^{(x)2}} - 
\frac{1}{W_{10}^{(y)2}} \right)
\sin \theta_0 \cos \theta_0. 
\nonumber
\end{eqnarray}    
Note that the initial condition for pulse 1 in the $(x,y,t)$ coordinate 
system is not separable, and therefore the current investigation also provides 
an example for collision-induced dynamics with a nonseparable 
initial condition.

Since the initial condition for pulse 1 is nonseparable, 
we calculate $\Delta A_{1}^{(c)}$ by using the general 
expression of Eq. (\ref{da17}). For the current setup, 
$C_{d1}=2\pi W_{10}^{(x)}W_{10}^{(y)}$. 
Additionally, the initial condition for pulse 2 is separable, 
and as a result, we can use Eq. (\ref{da31}), where $c_{d2}^{(x)}=(2\pi)^{1/2}$. 
Substituting these relations into Eq. (\ref{da17}), we obtain 
\begin{eqnarray} &&
\!\!\!\!
\Delta A_{1}^{(c)}=
-\frac{2^{1/2}\epsilon_{2}A_{1}(t_{c}^{-})A_{2}(t_{c}^{-})}{\pi^{1/2}|v_{d1}|}
\frac{W_{20}}{W_{10}^{(x)}W_{10}^{(y)}}
\!\int_{-\infty}^{\infty} \!\!\!\!\! dy \, g_{2}^{(y)}(y,t_{c})
\!\int_{-\infty}^{\infty} \!\!\!\!\! dx 
\;\tilde u_{10}(x,y,t_{c}). 
\label{da65}
\end{eqnarray}              
Since diffusion is isotropic, the unperturbed diffusion 
equation for pulse 1 in the $(x',y',t)$ coordinate system is: 
\begin{eqnarray}&&
\!\!\!\!\!\!\!
\partial _{t} \tilde u'_{10}=
\partial _{x'}^{2} \tilde u'_{10}+\partial _{y'}^{2} \tilde u'_{10}. 
\label{da66}
\end{eqnarray}  
Therefore, we can calculate $\tilde u_{10}(x,y,t)$ by solving 
Eq. (\ref{da66}) with the initial condition (\ref{da61}) in 
the $(x',y',t)$ coordinate system, and by expressing the solution 
in the $(x,y,t)$ coordinate system with the help of Eq. (\ref{da63}).  
Carrying out this calculation and multiplying the result by 
$g_{2}^{(y)}(y,t_{c})$, we obtain: 
\begin{eqnarray}&&
\!\!\!\!\!\!\!\!\!\!\!\!\!\!\!
g_{2}^{(y)}(y,t_{c})\tilde u_{10}(x,y,t_{c})=
\frac{W_{10}^{(x)}W_{10}^{(y)}W_{20}}
{(W_{10}^{(x)2} + 2t_{c})^{1/2}(W_{10}^{(y)2} + 2t_{c})^{1/2}
(W_{20}^{2} + 2t_{c})^{1/2}}
\nonumber \\&&
\times
\exp\left[-b_{1}^{2}x^2 - 2b_{2}xy - b_{3}^{2}y^2 \right],  
\label{da67}
\end{eqnarray}
where 
\begin{eqnarray}&&
b_{1} = \left( 
\frac{\cos^2 \theta_0}{2W_{10}^{(x)2} + 4t_{c}}
+ \frac{\sin^2 \theta_0}{2W_{10}^{(y)^{2}} + 4t_{c}}
\right)^{1/2},  
\label{da69}
\end{eqnarray}
\begin{eqnarray}&&
b_{2} = \left( 
\frac{1}{2W_{10}^{(x)2} + 4t_{c}}
- \frac{1}{2W_{10}^{(y)2} + 4t_{c}}
\right)\sin \theta_{0}\cos \theta_{0},
\label{da70}
\end{eqnarray}
\begin{eqnarray}&&
b_{3} = \left( 
\frac{1}{2W_{20}^{2} + 4t_{c}}
+ \frac{\sin^2 \theta_0}{2W_{10}^{(x)2} + 4t_{c}}
+ \frac{\cos^2 \theta_0}{2W_{10}^{(y)2} + 4t_{c}}
\right)^{1/2}.
\label{da71}
\end{eqnarray}
Substitution of Eq. (\ref{da67}) into Eq. (\ref{da65}) and 
integration with respect to $x$ and $y$ yield the following 
expression for $\Delta A_{1}^{(c)}$:
\begin{eqnarray}&&
 \!\!\!\!\!\!\!\!\!\!\!\!\!\!
\Delta A_{1}^{(c)} = \frac{-(2\pi)^{1/2}\epsilon_{2} A_{1}(t_{c}^{-})A_{2}(t_{c}^{-})}{|v_{d1}|}
\nonumber\\&&
\times
\frac{W_{20}^{2}}
{\left(W_{10}^{(x)2} + 2t_{c} \right)^{1/2}
\left(W_{10}^{(y)2} + 2t_{c} \right)^{1/2}
\left(W_{20}^{2} + 2t_{c} \right)^{1/2}
\left( b_{1}^{2}b_{3}^{2} - b_{2}^{2}\right)^{1/2} },
\!\!\!\!\!\!\!\!\!\!\!\!\!\!
\label{da72}
\end{eqnarray}
where 
\begin{eqnarray}&&
b_{1}^{2}b_{3}^{2} - b_{2}^{2}=
\frac{1}{2W_{20}^{2} + 4t_{c}}
\left( \frac{\cos^2 \theta_0}{2W_{10}^{(x)2} + 4t_{c}}
+ \frac{\sin^2 \theta_0}{2W_{10}^{(y)2} + 4t_{c}}\right)
\nonumber\\&&
+ \frac{1}{\left(2W_{10}^{(x)2} + 4t_{c}\right)
\left(2W_{10}^{(y)2} + 4t_{c}\right)}.
\label{da73}
\end{eqnarray}
Thus, even in the simple anisotropic collision setup 
considered here, the dependence of $\Delta A_{1}^{(c)}$ on the 
orientation angle $\theta_{0}$ is complex. This complex  
dependence of $\Delta A_{1}^{(c)}$ on $\theta_{0}$ can also be 
related to the nonseparable nature of the initial condition for pulse 1.

\begin{figure}[ptb]
\begin{center}
\epsfxsize=11cm \epsffile{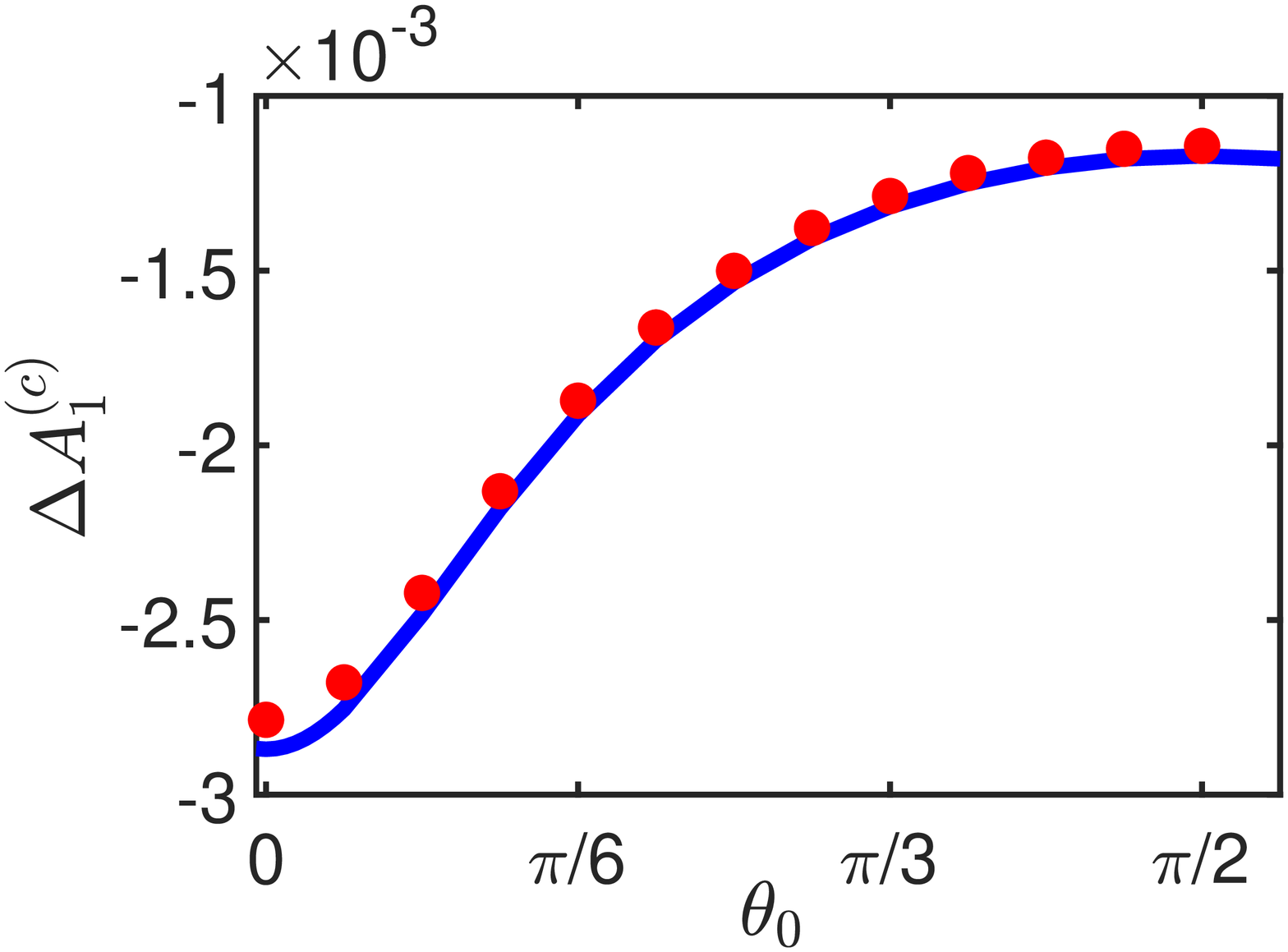}
\end{center}
\caption{(Color online) 
The collision-induced amplitude shift of pulse 1 $\Delta A_{1}^{(c)}$ 
vs the orientation angle $\theta_{0}$ in a fast collision between two 
Gaussian pulses with the anisotropic initial condition of 
Eqs. (\ref{da62}) and (\ref{da64}). The red circles represent 
the result obtained by numerical simulations with Eq. (\ref{da1}).  
The solid blue curve corresponds to the perturbation theory 
prediction of Eq. (\ref{da72}).} 
\label{fig6}
\end{figure}

We check the perturbation theory prediction of Eq. (\ref{da72}) for 
the dependence of $\Delta A_{1}^{(c)}$ on the orientation angle $\theta_{0}$ 
by numerical simulations with Eq. (\ref{da1}) with the initial condition of 
Eqs. (\ref{da62}) and (\ref{da64}). The simulations are carried out for 
$\theta_{0}$ values in the interval $0 \le \theta_{0} \le \pi/2$,  
and with physical parameter values $\epsilon_{2}=0.01$ and $v_{d1}=20$. 
The initial pulse parameters values are $A_{j}(0)=1$, $x_{20}=-20$, 
$W_{10}^{(x)}=8$, $W_{10}^{(y)}=2$, and $W_{20}=2$. The final time 
is $t_{f}=2$, and therefore, the pulses are well separated at $t_{f}$.   
The initial pulse shapes $u_{j}(x,y,0)$, and the pulse shapes $u_{j}(x,y,t)$ 
obtained in the simulation with $\theta_{0}=\pi/4$ at the intermediate 
time $t_{i}=1.2>t_{c}$, and at $t_{f}=2$ are shown in Fig. \ref{fig5}.  
We observe that both pulses undergo significant broadening due to 
diffusion, and that the maximum values of $u_{j}(x,y,t)$ decrease 
with increasing time. The dependence of $\Delta A_{1}^{(c)}$ 
on $\theta_{0}$ obtained in the simulations is shown in Fig. \ref{fig6} 
along with the theoretical prediction of Eq. (\ref{da72}). 
The agreement between the simulations result and the perturbation theory  
prediction is very good. More specifically, the relative error (in percentage) 
$|\Delta A_{1}^{(c)(num)}-\Delta A_{1}^{(c)(th)}|\times 100/|\Delta A_{1}^{(c)(th)}|$ 
is smaller than $2.9\%$ in the entire interval $0 \le \theta_{0} \le \pi/2$. 
Therefore, the numerical simulations validate the perturbation 
theory prediction for a complex dependence of $\Delta A_{1}^{(c)}$ on 
$\theta_{0}$ due to the anisotropic (and nonseparable) character 
of the initial condition.

\subsection{Collision-induced change in the pulse shape} 
\label{simu_reshaping}

The collision-induced change in the pulse shape in the transverse 
direction, which was uncovered by our perturbation theory in 
subsection \ref{da_general_IC}, is an important collisional effect 
that exists only in spatial dimension higher than 1. Indeed, it was 
shown in Refs. \cite{PNH2017B,NHP2021} that within the leading order 
of the perturbative calculation, the pulse shape is preserved 
during a fast two-pulse collision in the presence of weak quadratic loss 
in spatial dimension 1. In the current subsection, we investigate this important 
high-dimensional effect of the collision in detail both analytically 
and by numerical simulations.

To enable a more accurate comparison between the perturbative calculation  
and the numerical simulations, we assume that the effects of quadratic 
loss on single-pulse evolution, which are described by the 
terms proportional to $-u_{j}^{2}$ in Eq. (\ref{da1}), are negligible. 
This assumption is valid when the rates of the reactions $A + A \rightarrow A_{2}$ 
and $B + B \rightarrow B_{2}$ are much smaller than the rate of the reaction 
$A + B \rightarrow AB$. Furthermore, from Eq. (\ref{da10}) it follows that 
the terms $\tilde u_{j0}^{(1)}$, $\tilde u_{j0}^{(2)}$, etc. in the expansions 
(\ref{da7_add1}) of the $u_{j0}$ do not contribute to the leading-order expression 
for the collision-induced change in the pulse shape $\phi_{1}$. 
Therefore, the omission of the $-u_{j}^{2}$ terms in Eq. (\ref{da1}) is also 
mathematically consistent, since it does not affect the perturbation theory prediction 
and the numerical simulation result for the leading-order expression for $\phi_{1}$. 
Neglecting the terms proportional to $-u_{j}^{2}$ in Eq. (\ref{da1}), 
we obtain the following perturbed linear diffusion-advection equation for 
the collision dynamics:         
\begin{eqnarray} &&
\!\!\!\!\!\!\!\!\!\!\!\!\!\!
\partial _{t}u_{1}=\partial _{x}^{2}u_{1}+\partial _{y}^{2}u_{1}
-2\epsilon_{2}u_{2}u_{1},
\nonumber \\&&
\!\!\!\!\!\!\!\!\!\!\!\!\!\!
\partial _{t}u_{2}=\partial _{x}^{2}u_{2}+\partial _{y}^{2}u_{2}
-v_{d1}\partial _{x}u_{2} - 2\epsilon_{2}u_{1}u_{2}, 
\label{da81}
\end{eqnarray}   
where $0< \epsilon_{2} \ll 1$. We consider a collision between two 
Gaussian pulses as a concrete example. Therefore, the initial condition 
for the collision problem is given by Eq. (\ref{da51}). With this choice 
of the initial condition we can obtain an explicit formula for the 
collision-induced change in the shape of pulse 1 $\phi_{1}$ 
in the post-collision interval.

We note that the initial condition (\ref{da51}) is separable for 
both pulses. Therefore, we can calculate $\phi_{1}$ in the post-collision 
interval by using Eq. (\ref{da46}). Additionally, since the effects 
of quadratic loss on single-pulse evolution are negligible, 
we can replace $A_{j}(t_{c}^{-})$ by $A_{j}(0)$ everywhere 
in the calculation. Therefore, the coefficient $\tilde a_{1}$ 
of Eq. (\ref{da42}) is given by \cite{a1tc}:  
\begin{eqnarray} &&
\tilde a_{1}=
(8\pi)^{1/2}\epsilon_{2}A_{1}(0) A_{2}(0) W_{20}^{(x)}/|v_{d1}|.  
\label{da82}
\end{eqnarray}        
The function $g_{1}^{(x)}(x,t)$ in Eq. (\ref{da46}) is given by 
Eq. (\ref{appC_4}) in Appendix \ref{appendC}. In addition, 
in Appendix \ref{appendD}, we show that       
\begin{eqnarray}&&
{\cal F}^{-1}\left(\hat g_{12}^{(y)}(k_{2},t_{c})
\exp[- k_{2}^{2}(t-t_{c})]\right)= 
\frac{W_{10}^{(y)}W_{20}^{(y)}}
{(W_{10}^{(y)2} + 2t_{c})^{1/2}
(W_{20}^{(y)2} + 2t_{c})^{1/2}
\left[1+4\tilde a_{2}^{2}(t_{c})(t-t_{c})\right]^{1/2}}
\nonumber \\&&
\;\;\;\;\;\;\;\;\;
\times
\exp \left[-\frac{\tilde a_{2}^{2}(t_{c})y^{2}}{
1+4\tilde a_{2}^{2}(t_{c})(t-t_{c})} \right] , 
\label{da83}
\end{eqnarray} 
where $\tilde a_{2}^{2}(t_{c})$ is given by Eq. (\ref{appD_2}).   
Substitution of Eqs. (\ref{da82}), (\ref{da83}), and (\ref{appC_4}) into 
Eq. (\ref{da46}) yields the following expression for $\phi_{1}(x,y,t)$ 
in the post-collision interval: 
\begin{eqnarray}&&
\phi_{1}(x,y,t)= 
-\frac{\tilde a_{1} W_{10}^{(x)} W_{10}^{(y)} W_{20}^{(y)}}
{(W_{10}^{(x)2} + 2t)^{1/2} (W_{10}^{(y)2} + 2t_{c})^{1/2}
(W_{20}^{(y)2} + 2t_{c})^{1/2}
\left[1+4\tilde a_{2}^{2}(t_{c})(t-t_{c})\right]^{1/2}}
\nonumber \\&&
\;\;\;\;\;\;\;\;\;
\times
\exp \left[-\frac{x^{2}}{2W_{10}^{(x)2} + 4t} 
-\frac{\tilde a_{2}^{2}(t_{c})y^{2}}
{1+4\tilde a_{2}^{2}(t_{c})(t-t_{c})} \right].
\label{da85}
\end{eqnarray}

Further insight into the collision-induced change in the pulse 
shape can be gained by analyzing the fractional concentration 
reduction factor $\Delta \phi_{1}^{(r)}$. 
In the current collision setup, the effects of quadratic loss 
on single-pulse evolution are negligible. As a result, in this 
case, $u_{j0}(x,y,t) = A_{j}(0) \tilde u_{j0}(x,y,t)$, 
and from Eq. (\ref{da5}) we obtain:   
$u_{j}(x,y,t) = A_{j}(0) \tilde u_{j0}(x,y,t) + \phi_{j}(x,y,t)$.
Thus, in the current collision setup, we define 
$\Delta \phi_{1}^{(r)}$ by: 
\begin{eqnarray}&&
\Delta \phi_{1}^{(r)}(x,y,t) = 
\frac{A_{1}(0)\tilde u_{10}(x,y,t) - u_{1}(x,y,t)}{A_{1}(0)\tilde u_{10}(x,y,t)}
= -\frac{\phi_{1}(x,y,t)}{A_{1}(0)\tilde u_{10}(x,y,t)}.
\label{da86}
\end{eqnarray}  
We see that the fractional concentration reduction factor measures the ratio between 
the concentration decrease of pulse 1, which is induced by the effects 
of quadratic loss on the collision, and the concentration of pulse 1 in 
the unperturbed single-pulse evolution problem. 
Note that for a separable initial condition, the $x$ dependences of 
$\tilde u_{10}$ and of the leading order expression for $\phi_{1}$ are identical. 
Therefore, in this case, the $x$ dependence cancels out on 
the right hand side of Eq. (\ref{da86}), and $\Delta \phi_{1}^{(r)}$ 
becomes a function of $y$ and $t$ only.

\begin{figure}[ptb]
\begin{center}
\epsfxsize=11cm \epsffile{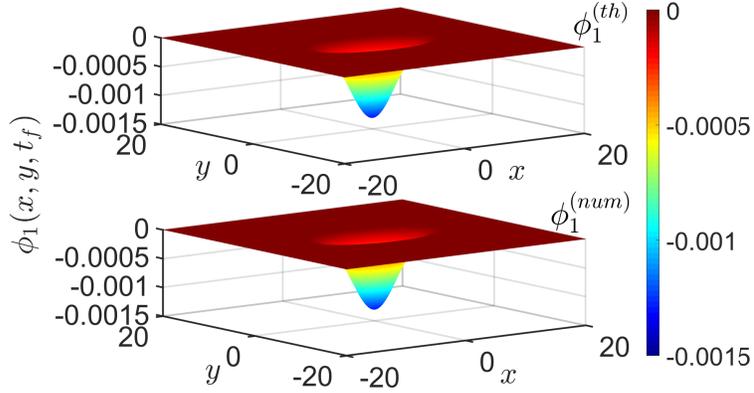}
\end{center}
\caption{(Color online) 
The collision-induced change in the shape of pulse 1 $\phi_{1}(x,y,t_{f})$ 
at $t_{f}=1.6$ in a fast two-pulse collision with physical parameter values  
$\epsilon_{2}=0.01$ and $v_{d1}=25$.   
Top: the perturbation theory prediction of Eq. (\ref{da85}). 
Bottom: the result obtained by numerical solution of Eq. (\ref{da81}).} 
\label{fig7}
\end{figure}

We check the perturbation theory predictions for the collision-induced 
change in the pulse shape by extensive numerical simulations 
with Eq. (\ref{da81}). We perform the simulations with $\epsilon_{2}=0.01$ 
and with $v_{d1}$ values satisfying $4 \le |v_{d1}| \le 60$. 
The parameter values of the initial condition (\ref{da51}) are $A_{j}(0)=1$,  
$x_{20}=\pm 20$, $W_{10}^{(x)}=3$, $W_{10}^{(y)}=2$, $W_{20}^{(x)}=2$, 
and $W_{20}^{(y)}=1$. The final time is $t_{f}=2t_{c}=-2x_{20}/v_{d1}$, 
and therefore, the pulses are well separated at $t_{f}$. 
The contour plots of the pulse shapes $u_{j}(x,y,t)$ obtained in the 
simulations are similar to the ones seen in Fig. \ref{fig1}.    
In particular, the pulses undergo significant diffusion-induced 
broadening, and the maximum values of $u_{j}(x,y,t)$ decrease with 
increasing $t$. The collision-induced change in the 
shape of pulse 1 that is obtained in the simulation with $v_{d1}=25$ 
at $t=t_{f}$, $\phi_{1}^{(num)}(x,y,t_{f})$, is shown in Fig. \ref{fig7}. 
The perturbation theory prediction, $\phi_{1}^{(th)}(x,y,t_{f})$, 
which is obtained by Eq. (\ref{da85}), is also shown. 
The agreement between the simulation result and the perturbation 
theory prediction is excellent. To quantify the deviation of 
$\phi_{1}^{(th)}(x,y,t)$ from  $\phi_{1}^{(num)}(x,y,t)$, we define  
the relative error (in percentage) $E_{r}^{(\phi_{1})}(t)$ by:     
\begin{eqnarray}&&
E_{r}^{(\phi_{1})}(t)=
100 \times
\left [\int dx\int dy \, |\phi_{1}^{(th)}(x,y,t)|^{2} \right ]^{-1/2}
\nonumber \\&&
\times
\left\{\int dx\int dy 
\left[\;\left|\phi_{1}^{(th)}(x,y,t) \right| - 
\left|\phi_{1}^{(num)}(x,y,t) \right| \; 
\right]^2 \right\}^{1/2},   
\label{da87}
\end{eqnarray}       
where the integration is performed over the entire simulation domain 
in the $xy$ plane. We find that the value of 
$E_{r}^{(\phi_{1})}(t_{f})$ is $0.82\%$ for $v_{d1}=25$, in 
accordance with the excellent agreement between simulation and 
theory observed in Fig. \ref{fig7}.

\begin{figure}[ptb]
\begin{center}
\epsfxsize=11cm  \epsffile{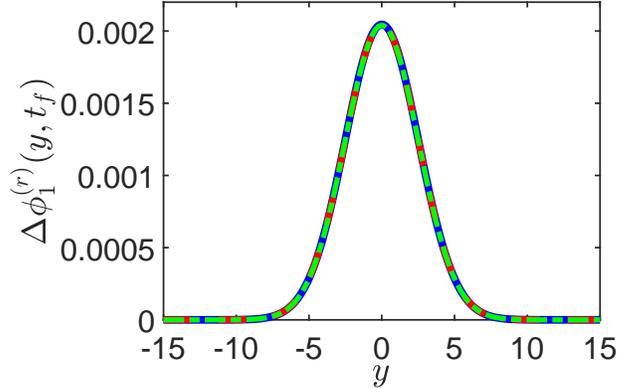} 
\end{center}
\caption{(Color online) 
The fractional concentration reduction factor for pulse 1  
at $t=t_{f}$, $\Delta \phi_{1}^{(r)}(y,t_{f})$, vs $y$   
in a two-pulse collision with parameter values $\epsilon_{2}=0.01$ and $v_{d1}=25$. 
The solid blue curve represents the perturbation theory prediction 
of Eqs. (\ref{da86}) and (\ref{da85}). The other two curves correspond to 
the results obtained by numerical solution of Eq. (\ref{da81}). 
The dashed red curve is obtained by averaging $\Delta \phi_{1}^{(r)}(x,y,t_{f})$  
over the $x$-interval $[-5,5]$. The dashed-dotted green curve is obtained by using 
the numerically computed value of $\Delta \phi_{1}^{(r)}(0,y,t_{f})$.}
\label{fig8}
\end{figure}

As a further check of the perturbation theory prediction, we analyze 
the $y$ dependence of the fractional concentration reduction factor $\Delta \phi_{1}^{(r)}$. 
Figure \ref{fig8} shows the $y$ dependence of $\Delta \phi_{1}^{(r)}(y,t_{f})$ 
obtained in the simulation with $v_{d1}=25$ \cite{Delta_phi_r}. 
A comparison with the perturbation theory prediction of Eqs. (\ref{da86}) 
and (\ref{da85}) is also shown. The agreement between the simulation  
result and the theoretical prediction is very good. 
The results in Figs. \ref{fig7} and \ref{fig8} and 
similar results that are obtained with other sets of the 
physical parameter values show that our perturbation method 
correctly captures the spatial distribution of the collision-induced 
change in the pulse shape.

\begin{figure}[ptb]
\begin{center}
\epsfxsize=11cm  \epsffile{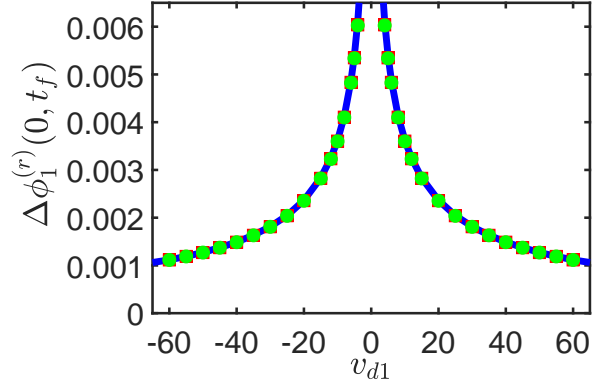} 
\end{center}
\caption{(Color online) 
The fractional concentration reduction factor for pulse 1 
at $y=0$ and $t=t_{f}$, $\Delta \phi_{1}^{(r)}(0,t_{f})$,  
vs advection velocity $v_{d1}$ in a fast two-pulse 
collision with $\epsilon_{2}=0.01$. The solid 
blue curve corresponds to the perturbation theory prediction 
of Eqs. (\ref{da86}) and (\ref{da85}). The other two curves are  
obtained from the numerical solution of Eq. (\ref{da81}). 
The red squares represent the result obtained by averaging 
$\Delta \phi_{1}^{(r)}(x,0,t_{f})$  over the $x$-interval $[-5,5]$. 
The green circles represent the result obtained by using the 
numerically computed value of $\Delta \phi_{1}^{(r)}(0,0,t_{f})$.}
\label{fig9}
\end{figure}

We now turn to study the dependence of the fractional concentration reduction factor 
on the advection velocity. We characterize this dependence by measuring  
$\Delta \phi_{1}^{(r)}(0,t_{f})$ as a function of $v_{d1}$. 
Figure \ref{fig9} shows the $v_{d1}$ dependence of $\Delta \phi_{1}^{(r)}(0,t_{f})$ 
obtained in the simulations along  with the theoretical prediction of 
Eqs. (\ref{da86}) and (\ref{da85}). 
The agreement between the simulations result and the perturbation theory 
prediction is excellent for all $v_{d1}$ values, $4 \le |v_{d1}| \le 60$. 
In particular, the relative error in the approximation of 
$\Delta \phi_{1}^{(r)}(0,t_{f})$, which is defined by 
$|\Delta \phi_{1}^{(r)(num)}(0,t_{f}) - \Delta \phi_{1}^{(r)(th)}(0,t_{f})| 
\times 100 / |\Delta \phi_{1}^{(r)(th)}(0,t_{f})|$, is smaller than 
$0.6\%$ for $ 10 \le |v_{d1}| \le 60$ and smaller than 
$1.4\%$ for $ 4 \le |v_{d1}| < 10$.
We also studied the dependence of $\Delta A_{1}^{(c)}$ on $v_{d1}$, 
and found very good agreement between the simulations result 
and the perturbation theory prediction (similar to what 
is shown in Figs. \ref{fig2} and \ref{fig4}). 
Based on these results and on the results shown in Figs. 
\ref{fig7} and \ref{fig8} we conclude that the numerical simulations 
confirm the third major prediction of our generalized perturbation 
theory for the collision-induced change in the pulse shape in 
the direction transverse to the advection velocity vector.

\section{Importance of collision-induced effects due to quadratic loss}
\label{discussion}

In the fast single-collision problem described by Eq. (\ref{da1}), 
the single-pulse evolution effects due to quadratic loss are 
of order $\epsilon_{2}$ and higher, while the collisional effects 
due to quadratic loss are of order $\epsilon_{2}/|v_{d1}|$ and higher. 
Thus, in this specific problem, the single-pulse evolution effects 
are stronger than the collision-induced effects. However, this does 
not reduce the importance of collision-induced effects due to quadratic 
loss in diffusion-advection systems, since there are many physical setups 
where the collisional effects can be comparable to or stronger 
than the single-pulse evolution effects. 
One can distinguish between two types of physical systems, for which the 
latter statement is valid. In the first type, which is described 
in subsection \ref{simu_reshaping}, the rates of the reactions $A + A \rightarrow A_{2}$ 
and $B + B \rightarrow B_{2}$ are much smaller than the rate of the reaction 
$A + B \rightarrow AB$. As a result, in these systems, the collision-induced 
effects due to quadratic loss are stronger than the corresponding 
single-pulse evolution effects already in the single-collision problem. 
In the second type of systems, the collisional effects due to quadratic loss  
are comparable to or stronger than the corresponding single-pulse evolution effects, 
even though the latter effects are stronger than the former in a single collision. 
This is the situation in physical systems containing a large number of pulses 
of different substances that are moving with different velocities, 
since the high collision rates in these systems can lead to dominance of 
the collisional effects compared with single-pulse evolution effects.   
An important example for this situation is provided by multisequence systems 
(including multisequence chemical communication links), 
where multiple pulse sequences move through the same medium with different velocities. 
In the current section, we briefly discuss the importance of collision-induced effects 
due to quadratic loss in comparison with single-pulse evolution effects due 
to quadratic loss in multisequence systems.

We start by considering two periodic pulse sequences of substances 1 and 2 (two gases), 
which are moving with different velocities through the same gaseous medium. We assume 
that the vector of relative velocity between the pulse sequences lies on the 
$x$ axis, and denote the magnitude of this vector by $|v_{d1}|$. In addition, 
we denote the spatial separation (along the $x$ axis) between the centers of any 
two neighboring pulses in the two sequences by $b$. Then, the rate of collisions 
between a pulse from sequence 1 and pulses from sequence 2 is $|v_{d1}|/b$. As a result, 
the rate of change of the amplitude and shape of a pulse from sequence 1 due to 
collisions with sequence 2 pulses in the presence of weak quadratic loss is 
proportional to:        
\begin{equation} 
\left(|v_{d1}|/b \right) \times \left(-\epsilon_{2}/ |v_{d1}|\right) = -\epsilon_{2}/b . 
\nonumber 
\end{equation}         
Therefore, the magnitude of the collisional effects is comparable to the magnitude of 
single-pulse evolution effects already in a two-sequence system. 
Furthermore, by a similar argument, in a system with $N$ periodic pulse sequences 
of $N$ different reacting substances, the rate of change of the amplitude and shape of a given pulse 
due to the cumulative effects of quadratic loss on the collisions is proportional to $-(N-1)\epsilon_{2}/b$. 
As a result, when $N \gg 1$, the collision-induced effects due to quadratic loss are dominant 
compared with the single-pulse evolution effects due to quadratic loss.

We emphasize that predictions about the importance of collision-induced effects 
in comparison with single-pulse evolution effects due to nonlinear dissipation 
in multisequence systems have been confirmed by extensive numerical simulations 
for propagation of multiple periodic sequences of optical solitons through 
an optical waveguide \cite{PNC2010,PC2012,PNT2016,PNH2017A,PC2018}. 
In this case, the dynamics is described by a system of perturbed nonlinear 
Schr\"odinger equations, where the dissipative perturbations are due to nonlinear 
loss and / or delayed Raman response \cite{PNC2010,PC2012,PNT2016,PNH2017A,PC2018}.   
The collision rate calculations for periodic sequences of pulses of the linear diffusion-advection 
system are essentially the same as the one used for multiple periodic soliton sequences 
in Refs. \cite{PNC2010,PC2012,PNT2016,PNH2017A,PC2018}. 
Therefore, the same conclusion is expected to hold for periodic sequences of the pulses considered 
in the current paper. That is, the collision-induced effects due to quadratic loss in the 
multisequence system can be comparable to or stronger than the single-pulse evolution effects 
due to quadratic loss, even though the latter effects are stronger than the former 
in a single collision.

\section{Conclusions}
\label{conclusions}

We investigated the dynamics of fast two-pulse collisions 
in linear diffusion-advection systems with weak quadratic loss 
in spatial dimension 2. The quadratic loss arises due to 
chemical reactions. We first introduced a two-dimensional 
perturbation method, which generalizes the perturbation method 
used in Refs. \cite{PNH2017B,NHP2021} for studying the one-dimensional 
collision problem in three important aspects. 
First, it extends the perturbative calculation from spatial 
dimension 1 to spatial dimension 2, and enables the extension 
of the calculation to a general spatial dimension. 
Second, it provides a calculation of the collision-induced 
dynamics of the pulse shape both in the collision interval and 
outside of the collision interval, whereas the calculation 
in Refs. \cite{PNH2017B,NHP2021} was limited to the collision 
interval only. In this way, our generalized perturbation method enables for the 
first time an accurate comparison between perturbation theory predictions 
and numerical simulations results for the collision-induced 
change in the pulse shape. Third, it enables the discovery 
and analysis of several major collision-induced effects, 
which exist only in the high-dimensional problem. 
A key ingredient in the generalization of the perturbation 
method is the application of a rotation transformation, 
such that in the new coordinate system, the advection velocity 
vector lies on the $x$-axis. 
The application of this transformation enables one to overcome 
the main challenges in generalizing the perturbation method to 
spatial dimension higher than 1. More specifically, it enables one 
to preserve the true small parameters in the problem. Moreover, it simplifies 
the calculation, and in this manner, it enables the derivation of 
explicit expressions for the collision-induced changes in the 
pulse shape and amplitude.

We used the generalized perturbation method to obtain 
formulas for the collision-induced changes in the pulse shapes 
and amplitudes in spatial dimension 2. 
We showed that for a general initial condition, the collision 
induces a change in the pulse shape in the direction transverse 
to the advection velocity vector. We also considered the important 
case of a separable initial condition, and showed that in this 
case, the pulse shape in the longitudinal direction is not 
changed by the collision in the leading order of the perturbation 
theory. Furthermore, we found that for a separable initial condition, 
the longitudinal part in the expression for the collision-induced 
amplitude shift has a simple universal form. This finding is important,  
since it can be used in the design of scalable multisequence 
chemical communication links. Additionally, the transverse part in the 
expression for the amplitude shift was found to be nonuniversal, 
and proportional to the integral of the product of the pulse 
concentrations with respect to the transverse coordinate.  
We also showed that anisotropy in the initial condition 
leads to a complex dependence of the expression for the collision-induced 
amplitude shift on the orientation angle between the pulses.     
This complex dependence was attributed to the nonseparable 
nature of the initial condition in the anisotropic case.

We obtained very good agreement between our perturbation theory predictions 
and the results of extensive numerical simulations with the weakly 
perturbed diffusion-advection model for all the high-dimensional effects 
described in the preceding paragraph. Therefore, our work significantly 
enhanced and generalized the results of the previous studies in Refs. 
\cite{PNH2017B,NHP2021} on fast two-pulse collisions in linear 
diffusion-advection systems with weak quadratic loss, which were 
limited to spatial dimension 1. Interestingly, we have recently found that 
similar high-dimensional effects exist in fast two-beam collisions 
in bulk linear optical media with weak cubic loss \cite{PHN2021}. 
We emphasize that detailed analytic results on collisions 
between pulse solutions of linear or nonlinear evolution models 
in the presence of nonlinear dissipation in spatial dimension 
higher than 1 are scarce, and this is especially true for pulses 
that are not shape preserving. Therefore, the current study and 
our recent study in Ref. \cite{PHN2021} also significantly extended 
the understanding of the general high-dimensional problem 
of fast two-pulse collisions in the presence of nonlinear dissipation.

\appendix
\section{Invariance of $\Delta A_{1}^{(c)}$ under rotation transformations} 
\label{appendA}

In the current Appendix, we show that the change in the coordinate system, 
in which we rotate the $x'$ and $y'$ axes by an angle $\Delta\theta$, 
such that in the new coordinate system the advection velocity vector is on 
the $x$ axis, does not change the value of $\Delta A_{1}^{(c)}$. 
That is, the value of $\Delta A_{1}^{(c)}$ is invariant under rotation 
transformations in the $xy$ plane. The change in the coordinate system 
and the associated invariance property are important for the 
following reasons. First, only the formulas obtained in the new 
coordinate system explicitly preserve the true small parameters in the problem. 
Second, the simpler form of the latter formulas enables the derivation of 
explicit expressions for $\Delta A_{1}^{(c)}$.               
In this manner, the application of the rotation transformation provides deeper 
insight into the collision dynamics in spatial dimension higher than 1.

Consider the fast collision problem in the $(x',y',t)$ 
coordinate system, in which the advection velocity vector 
$\mathbf{v_{d}'}=(v_{d1}', v_{d2}')$ does not lie on 
the $x'$ or $y'$ axes. We assume that $v_{d}'=|\mathbf{v_{d}'}| \gg 1$. 
Therefore, the two small parameters in the problem are $\epsilon_{2}$ 
and $1/v_{d}'$. The perturbed linear diffusion-advection model in the 
$(x',y',t)$ coordinate system is        
\begin{eqnarray} &&
\!\!\!\!\!\!\!\!\!\!\!\!\!\!
\partial _{t}u'_{1}=\partial _{x'}^{2}u'_{1}+\partial _{y'}^{2}u'_{1}
-\epsilon_{2}u_{1}^{'2}-2\epsilon_{2}u'_{2}u'_{1},
\nonumber \\&&
\!\!\!\!\!\!\!\!\!\!\!\!\!\!
\partial _{t}u'_{2}=\partial_{x'}^{2}u'_{2}+\partial _{y'}^{2}u'_{2}
- v'_{d1}\partial _{x'}u'_{2} - v'_{d2}\partial _{y'}u'_{2}  
-\epsilon_{2}u_{2}^{'2}-2\epsilon_{2}u'_{1}u'_{2} , 
\label{appA_1}
\end{eqnarray}    
where $u'_{j}(x',y',t)$ is the concentration of substance $j$ in this 
coordinate system. The initial condition for the collision 
problem is: 
\begin{eqnarray} &&
u'_{j}(x',y',0)=A_{j}(0)h'_{j}(x',y').  
\label{appA_2}
\end{eqnarray}

We assume that the solution to the unperturbed diffusion equation 
\begin{eqnarray}&&
\!\!\!\!\!\!\!
\partial _{t}u'_{1}=\partial _{x'}^{2}u'_{1}+\partial _{y'}^{2}u'_{1}
\label{appA_3}
\end{eqnarray} 
does not contain any fast dependence on $t$. We also assume 
that the only fast dependence on $t$ in the solution 
to the equation 
\begin{eqnarray}&&
\!\!\!\!\!\!\!  
\partial _{t}u'_{2}=\partial_{x'}^{2}u'_{2}+\partial _{y'}^{2}u'_{2}
- v'_{d1}\partial _{x'}u'_{2} - v'_{d2}\partial _{y'}u'_{2}  
\label{appA_4}
\end{eqnarray}                      
is contained in factors of the form $x'-x'_{20}-v_{d1}'t$ 
and $y'-y'_{20}-v_{d2}'t$, where $(x'_{20},y'_{20})$ is the 
initial position of pulse 2 in the $x'y'$ plane. 
When these assumptions are satisfied, we can use the perturbation method 
of subsection \ref{da_general_IC} to show that the equation for $\phi_{1}'$ 
in the collision interval in the leading order of the calculation is   
\begin{equation}
\partial _{t}\phi'_{1}=-2\epsilon_{2}A_{1}(t)A_{2}(t) \tilde u'_{20} \tilde u'_{10}.
\label{appA_5} 
\end{equation}            
In addition, we can show by a similar calculation to the one 
carried out in subsection \ref{da_general_IC} that 
$\Delta\phi'_{1}(x',y',t_{c})$ can be approximated by: 
\begin{eqnarray}&&
\!\!\!\!\!\!\!
\Delta\phi'_{1}(x',y',t_{c})\!=\!
-2\epsilon_{2} A_{1}(t_{c}^{-}) A_{2}(t_{c}^{-})
\tilde u_{10}^{'}(x',y',t_{c})
\nonumber \\&&
\!\!\! \times
\int_{-\infty}^{\infty} \!\!\!\!\! dt' \, 
\bar u_{20}^{\prime}(x'-x'_{20}-v_{d1}'t',y'-y'_{20}-v_{d2}'t',t_{c}).
\label{appA_6}
\end{eqnarray}                
As a result, the collision-induced amplitude shift in the $(x',y',t)$ 
coordinate system is 
\begin{eqnarray} &&
\!\!\!\!\!\!\!\!
\Delta A_{1}^{'(c)}=-\frac{2\epsilon_{2}A_{1}(t_{c}^{-})A_{2}(t_{c}^{-})}{C'_{d1}}
\nonumber \\&&
\!\!\! \times
\!\!\int_{-\infty}^{\infty} \!\!\!\!\! dx' 
\!\int_{-\infty}^{\infty} \!\!\!\!\! dy'
\;\tilde u'_{10}(x',y',t_{c})
\int_{-\infty}^{\infty} \!\!\!\!\! dt' \, 
\bar u_{20}^{\prime}(x'-x'_{20}-v_{d1}'t',y'-y'_{20}-v_{d2}'t',t_{c}), 
\label{appA_7}
\end{eqnarray}  
where 
\begin{eqnarray}&&
C'_{d1}= 
\!\!\int_{-\infty}^{\infty} \!\!\!\!\! dx' 
\!\int_{-\infty}^{\infty} \!\!\!\!\! dy'
\;\tilde u_{10}^{\prime}(x',y',0).  
\label{appA_8} 
\end{eqnarray}       
Equation (\ref{appA_7}) does not explicitly preserve the 
true small parameters in the problem $\epsilon_{2}$ and 
$1/v_{d}'$, since it contains {\it two} parameters 
$v_{d1}'$ and $v_{d2}'$ that are associated with $v_{d}'$.     
Moreover, the integration variable $t'$ appears in the expression 
for $\bar u_{20}^{\prime}$ in the inner integral twice,  
and as a result, derivation of explicit formulas for the 
amplitude shift from Eq. (\ref{appA_7}) is challenging.

We now apply a rotation transformation from the $(x',y',t)$  
coordinate system to the $(x,y,t)$ coordinate system, in which 
the advection  velocity vector is on the $x$ axis. The rotation 
angle is $\Delta\theta = \arctan(v_{d2}'/v_{d1}')$, and the 
equations that define the transformation are:  
\begin{eqnarray}&&
x' = x \cos\Delta\theta - y\sin\Delta\theta,
\nonumber \\&&
y' = x \sin\Delta\theta + y\cos\Delta\theta, 
\label{appA_9} 
\end{eqnarray}  
and 
\begin{equation}
u'_{j}(x',y',t) = u_{j}(x,y,t).  
\label{appA_10} 
\end{equation}                     
One can show that the perturbed linear diffusion-advection model in the 
$(x,y,t)$ system is Eq. (\ref{da1}), where $|v_{d1}|=v_{d}'$.  
The initial condition for the two-pulse collision problem is given by 
Eq. (\ref{da2}), where $h_{j}(x,y) = h'_{j}(x',y')$. 
Since the only large parameter in Eq. (\ref{da1}) 
is $|v_{d1}|$, the transformation in Eqs. (\ref{appA_9}) and (\ref{appA_10}) 
does not change the properties of the fast dependence on $t$ of the solutions 
to the unperturbed diffusion equations (\ref{appA_3}) and (\ref{appA_4}). 
That is, the solution to the unperturbed equation 
\begin{eqnarray}&&
\!\!\!\!\!\!\!
\partial_{t}u_{1}=\partial_{x}^{2}u_{1} + \partial_{y}^{2}u_{1} 
\!\!\!\!\!\!\!\!
\label{appA_13} 
\end{eqnarray}          
does not contain any fast dependence on $t$, and the only 
fast dependence on $t$ in the solution to the equation
\begin{eqnarray}&&
\!\!\!\!\!\!\!
\partial _{t}u_{2}=\partial _{x}^{2}u_{2} + \partial_{y}^{2}u_{2}
-v_{d1}\partial _{x}u_{2} 
\!\!\!\!\!\!\!\!
\label{appA_14} 
\end{eqnarray}  
is contained in factors of the form $x-x_{20}-v_{d1}t$. 
It follows that we can calculate the collision-induced 
amplitude shift $\Delta A_{1}^{(c)}$ in the $(x,y,t)$ system 
by the perturbation method of subsection \ref{da_general_IC}. 
Moreover, the result of this calculation is that $\Delta A_{1}^{(c)}$ 
is given by Eqs. (\ref{da17}) and (\ref{da16}).

We now show that the amplitude shift value $\Delta A_{1}^{\prime (c)}$ 
in Eq. (\ref{appA_7}) is equal to the value of $\Delta A_{1}^{(c)}$ in Eq. (\ref{da17}).  
We first note that the Jacobian of the transformation (\ref{appA_9}) is equal to $1$.  
Using this together with Eqs. (\ref{da16}), (\ref{appA_8}), and (\ref{appA_10}), 
we find $C'_{d1}=C_{d1}$. Second, from Eq. (\ref{appA_10}) it follows 
that $\tilde u'_{j0}(x',y',t_{c}) = \tilde u_{j0}(x,y,t_{c})$.        
Third, since the transformation in Eqs. (\ref{appA_9})-(\ref{appA_10}) 
does not change the properties of the fast dependence on $t$ of the solutions to 
the unperturbed diffusion equations, and since $v_{d2}=0$, we obtain \cite{yy_20}:  
\begin{equation}
\bar u'_{20}(x'-x'_{20}-v_{d1}'t,y'-y'_{20}-v_{d2}'t,t_{c}) = 
\bar u_{20}(x-x_{20}-v_{d1}t,y,t_{c}). 
\label{appA_15} 
\end{equation}                        
Substituting all these relations into Eq. (\ref{appA_7}), we obtain: 
\begin{eqnarray} &&
\!\!\!\!\!\!\!\!
\Delta A_{1}^{\prime (c)}=-\frac{2\epsilon_{2}A_{1}(t_{c}^{-})A_{2}(t_{c}^{-})}
{C_{d1}|v_{d1}|}
\!\!\int_{-\infty}^{\infty} \!\!\!\!\! dx 
\!\int_{-\infty}^{\infty} \!\!\!\!\! dy
\;\tilde u_{10}(x,y,t_{c})
\!\int_{-\infty}^{\infty} \!\!\!\!\! d\tilde x \;\bar u_{20}(\tilde x,y,t_{c})
=\Delta A_{1}^{(c)}. 
\label{appA_16} 
\end{eqnarray}
Therefore, the value of $\Delta A_{1}^{(c)}$ is invariant 
under rotation transformations in the $xy$ plane.

\section{Amplitude dynamics in the perturbed single-pulse evolution problem} 
\label{appendB}

In this Appendix, we obtain the equation for the dynamics 
of pulse amplitudes in the perturbed single-pulse evolution 
problem. This equation is used in the calculation of the 
amplitude values in the approximate expressions (\ref{da10}) for the $u_{j0}$, 
and also in the calculation of $A_{j}(t_{c}^{-})$ in the 
equations for $\Delta\phi_{1}(x,y,t_{c})$ and 
$\Delta A_{1}^{(c)}$ in sections \ref{da_2D} and \ref{simu}.

We consider the dynamic evolution of a single pulse in the 
presence of diffusion, advection, and weak quadratic loss. 
The dynamics is described by Eqs. (\ref{da6}) and (\ref{da7}) 
for pulses 1 and 2, respectively. Using mass balance calculations 
for these  equations, we find  
\begin{eqnarray}&&
\!\!\!\!\!\!\!\!\!\!\!\!
\partial_{t}\int_{-\infty}^{\infty} \!\!\!\!\! dx 
\int_{-\infty}^{\infty} \!\!\!\!\! dy \, u_{j0}(x,y,t) \!=\!
-\epsilon_{2}\int_{-\infty}^{\infty} \!\!\!\!\! dx 
\int_{-\infty}^{\infty} \!\!\!\!\! dy \, u_{j0}^{2}(x,y,t).
\label{appB_1}
\end{eqnarray}       
We substitute the approximate expressions (\ref{da10}) for 
the $u_{j0}$ into Eq. (\ref{appB_1}), and obtain the following 
equation for the dynamics of the $A_{j}$: 
\begin{eqnarray}&&
\!\!\!\!\!\!\!\!\!\!\!\!
C_{dj}\frac{dA_{j}}{dt}= 
-\epsilon_{2}P_{2j}(t)A_{j}^{2} \,.  
\label{appB_2}
\end{eqnarray}      
In Eq. (\ref{appB_2}), $P_{2j}(t)=\int_{-\infty}^{\infty} \!\!\!\!\! dx 
\int_{-\infty}^{\infty} \!\!\!\!\! dy \,\tilde u_{j0}^{2}(x,y,t)$, 
$C_{d1}$ is given by Eq. (\ref{da16}), and $C_{d2}$ is given by 
a similar equation, in which $\tilde u_{10}(x,y,0)$ is replaced 
by $\tilde u_{20}(x,y,0)$. The solution of Eq. (\ref{appB_2}) 
on the time interval $[0,t]$ is 
\begin{eqnarray}&&
\!\!\!\!\!\!\!\!\!\!\!\!
A_{j}(t)=\frac{A_{j}(0)}
{1+\epsilon_{2}\tilde P_{2j}(0,t)A_{j}(0)/C_{dj}} \,,   
\label{appB_3}
\end{eqnarray}  
where $\tilde P_{2j}(0,t)=\int_{0}^{t} dt'\, P_{2j}(t')$.  
The effects of linear loss on the dynamics of pulse amplitudes  
can be included in the perturbative calculation in the same manner 
as was done in Refs. \cite{PNH2017B,NHP2021} for the one-dimensional 
problem. Furthermore, similar to the one-dimensional problem, it can 
be shown that these effects do not alter the form of the expressions for 
the collision-induced changes in pulse amplitudes and shapes.

\section{The solution of the unperturbed diffusion equation with a Gaussian initial condition} 
\label{appendC}

We present here a very brief summary of the formulas for the solution 
of the unperturbed linear diffusion equation with a Gaussian initial 
condition, since this solution is used extensively in section 
\ref{simu}, as an example. We consider the unperturbed linear diffusion equation 
\begin{eqnarray}&&
\!\!\!\!\!\!\!
\partial_{t}u=\partial_{x}^{2}u + \partial_{y}^{2}u 
\!\!\!\!\!\!\!\!
\label{appC_1} 
\end{eqnarray}          
with the separable Gaussian initial condition 
\begin{eqnarray}&&
u(x,y,0)=A\exp \left[-\frac{(x-x_{0})^{2}}{2W^{(x)2}_{0}}
-\frac{(y-y_{0})^{2}}{2W^{(y)2}_{0}} \right]. 
\label{appC_2}
\end{eqnarray}  
The solution of Eq. (\ref{appC_1}) with the initial 
condition (\ref{appC_2}) can be written as: 
\begin{eqnarray}&&
u(x,y,t)=Ag^{(x)}(\tilde x,t)g^{(y)}(\tilde y,t),
\label{appC_3}
\end{eqnarray}    
where $\tilde x = x - x_{0}$, $\tilde y = y - y_{0}$,          
\begin{eqnarray}&&
\!\!\!\!\!\!\!\!\!\!\!\!\!\!
g^{(x)}(\tilde x,t)= 
\frac{W_{0}^{(x)}}{(W_{0}^{(x)2} + 2t)^{1/2}}
\exp\left[-\frac{\tilde x^{2}}{2W_{0}^{(x)2} + 4t} \right], 
\label{appC_4}
\end{eqnarray}
and 
\begin{eqnarray}&&
\!\!\!\!\!\!\!\!\!\!\!\!\!\!
g^{(y)}(\tilde y,t)= 
\frac{W_{0}^{(y)}}{(W_{0}^{(y)2} + 2t)^{1/2}}
\exp\left[-\frac{\tilde y^{2}}{2W_{0}^{(y)2} + 4t} \right].
\label{appC_5}
\end{eqnarray}
In addition, the solution of Eq. (\ref{appC_1}) with the term 
$-v_{d1}\partial_{x} u$ on its right hand side and with the initial 
condition (\ref{appC_2}) is given by Eqs. (\ref{appC_3})-(\ref{appC_5})  
with $\tilde x = x - x_{0} - v_{d1}t$, and $\tilde y = y - y_{0}$.

\section{Derivation of Eq. (\ref{da83})}
\label{appendD}

We present the derivation of Eq. (\ref{da83}) for the inverse Fourier transform of  
$\hat g_{12}^{(y)}(k_{2},t_{c})\exp[-k_{2}^{2}(t-t_{c})]$. This equation is used 
in subsection \ref{simu_reshaping} in the calculation of $\phi_{1}(x,y,t)$ in 
the post-collision interval, in the case where the initial condition for 
the collision problem is given by the Gaussian pulses of Eq. (\ref{da51}).

Using Eqs. (\ref{da43}) and (\ref{appC_5}), we obtain: 
\begin{eqnarray}&&
g_{12}^{(y)}(y,t_{c})=
\frac{W_{10}^{(y)}W_{20}^{(y)}
\exp\left[-\tilde a_{2}^{2}(t_{c})y^{2}\right]}
{(W_{10}^{(y)2} + 2t_{c})^{1/2}
(W_{20}^{(y)2} + 2t_{c})^{1/2}} , 
\label{appD_1}     
\end{eqnarray}  
where 
\begin{equation}
\tilde a_{2}^{2}(t_{c})= 
\frac{W_{10}^{(y)2} + W_{20}^{(y)2} + 4t_{c}}
{2(W_{10}^{(y)2} + 2t_{c})(W_{20}^{(y)2} + 2t_{c})}. 
\label{appD_2}
\end{equation}     
The Fourier transform of $g_{12}^{(y)}(y,t_{c})$ is: 
\begin{eqnarray}&&
\hat g_{12}^{(y)}(k_{2},t_{c})=
\frac{W_{10}^{(y)}W_{20}^{(y)}}
{(W_{10}^{(y)2} + W_{20}^{(y)2} + 4t_{c})^{1/2}} 
\exp\left[ -\frac{k_{2}^{2}}{4\tilde a_{2}^{2}(t_{c})}\right] .  
\label{appD_3}     
\end{eqnarray}  
As a result, we obtain the following expression for the inverse Fourier 
transform of $\hat g_{12}^{(y)}(k_{2},t_{c})\exp[-k_{2}^{2}(t-t_{c})]$: 
\begin{eqnarray}&&
{\cal F}^{-1}\left(\hat g_{12}^{(y)}(k_{2},t_{c})
\exp[- k_{2}^{2}(t-t_{c})]\right)= 
\frac{W_{10}^{(y)}W_{20}^{(y)}}
{(W_{10}^{(y)2} + 2t_{c})^{1/2}
(W_{20}^{(y)2} + 2t_{c})^{1/2}
\left[1+4\tilde a_{2}^{2}(t_{c})(t-t_{c})\right]^{1/2}}
\nonumber \\&&
\;\;\;\;\;\;\;\;\;
\times
\exp \left[-\frac{\tilde a_{2}^{2}(t_{c})y^{2}}{
1+4\tilde a_{2}^{2}(t_{c})(t-t_{c})} \right]. 
\label{appD_4}
\end{eqnarray}          
Equation (\ref{appD_4}) is Eq. (\ref{da83}) of 
subsection \ref{simu_reshaping}.


\begin{thebibliography}{}
\bibitem{Dimensions1} The dimensionless coordinates $x$ and $y$ in 
Eq. (\ref{da1}) are $x=X/x_{0}$ and $y=Y/x_{0}$, where $X$ and $Y$ 
are the dimensional coordinates, and $x_{0}$ is the width of a 
reference pulse in the $x$ direction.  
The dimensionless time is $t=\tau/\tau_{D}$, where $\tau$ is time, 
$\tau_{D}=x_{0}^{2}/D$, and $D$ is the dimensional diffusion coefficient.     
$u_{j}=U_{j}/\rho_{0}$, where $U_{j}$ is the 
concentration of substance $j$ and $\rho_{0}$ is the peak concentration.  
$v_{d1}=x_{0}V_{d1}/D$, where $V_{d1}$ is the dimensional advection 
velocity (the dimensional velocity difference between the two pulses).  
$\epsilon_{2}=\rho_{0}x_{0}^{2}\tilde\epsilon_{2}/D$, where 
$\tilde\epsilon_{2}$ is the  dimensional quadratic loss coefficient.
\bibitem{ti_values} The values of $t_{i}$ are determined by:  
$t_{i}=t_{c}+r(t_{f}-t_{c})$, where $r=1/5$, as an example. 
Therefore, $t_{i}$ is an intermediate time that is larger 
than $t_{c}$, at which the collision is not yet completed.
\bibitem{a1tc} In subsection \ref{simu_reshaping}, the coefficient $\tilde a_{1}$ 
depends on $A_{1}(0)$ and $A_{2}(0)$ instead of on $A_{1}(t_{c}^{-})$ and 
$A_{2}(t_{c}^{-})$. Therefore, we replace the notation 
$\tilde a_{1}(t_{c}^{-})$ by $\tilde a_{1}$ in this subsection.
\bibitem{Delta_phi_r} Since the fractional concentration reduction factor obtained in 
the simulation $\Delta \phi_{1}^{(r)(num)}(x,y,t)$ shows weak dependence 
on $x$, we use two different methods to obtain $\Delta \phi_{1}^{(r)}(y,t)$ 
from the simulation result. In the first method, 
we calculate $\Delta \phi_{1}^{(r)}(y,t)$  by averaging 
$\Delta \phi_{1}^{(r)(num)}(x,y,t)$ over the $x$-interval $[-5,5]$,  
and in the second method, we use the value of 
$\Delta \phi_{1}^{(r)(num)}(0,y,t)$.
\bibitem{yy_20} Note that $\bar u_{20}$ depends on $y_{20}$. 
However, for brevity of notation, we did not write this dependence 
explicitly in Eq. (\ref{appA_15}) and throughout the paper.
\end{thebibliography}
\end{document}